\documentclass[manuscript,screen]{acmart}
\usepackage{multicol}
\usepackage{multirow}
\usepackage{soul}
\usepackage{lipsum}
\usepackage{wrapfig}
\usepackage{subfigure}
\usepackage[edges]{forest}
\usepackage{varwidth}

\newcommand{\subsubsubsection}[1]{\paragraph{#1}}
\expandafter\newcommand\csname r@tocindent4\endcsname{4in}

\usepackage[usestackEOL]{stackengine}
\edef\tmp{\the\baselineskip}
\setstackgap{L}{\tmp}

\AtBeginDocument{%
  \providecommand\BibTeX{{%
    \normalfont B\kern-0.5em{\scshape i\kern-0.25em b}\kern-0.8em\TeX}}}

\setcopyright{acmcopyright}
\copyrightyear{2022}
\acmDOI{10.1145/3652953}

\acmBooktitle{Woodstock '18: ACM Symposium on Neural Gaze Detection,
  June 03--05, 2018, Woodstock, NY}
\acmPrice{15.00}
\acmISBN{978-1-4503-XXXX-X/18/06}

\usepackage{glossaries}
\newacronym{cps}{CPS}{Cyber-Physical Systems}
\newacronym{cpss}{CPS}{Cyber-Physical Systems}
\newacronym{scada}{SCADA}{Supervisory Control and Data Acquisition}
\newacronym{mtd}{MTD}{Moving Target Defense}

\newcommand{\etal}{\textit{et al.}~}
\newcommand{\ie}{i.e.}

\newcommand{\blueremark}[2]{
  \textcolor{black}{#2}
}
\newcommand\CH[1]{\blueremark{}{#1}}
\usepackage{color}

\begin{document}

\title{\CH{A Survey on Cyber-Resilience Approaches for Cyber-Physical Systems}}

\author{Mariana Segovia-Ferreira}
\authornotemark[1]
%\authornote{~~}
\email{segovia@telecom-sudparis.eu}
\author{Jose Rubio-Hernan}
\authornotemark[1]
\email{rubio_he@telecom-sudparis.eu}
\author{Ana Rosa Cavalli}
\authornotemark[1]
\email{ana.cavalli@telecom-sudparis.eu}
\author{Joaquin Garcia-Alfaro}
\authornotemark[1]
\email{joaquin.garcia\_alfaro@telecom-sudparis.eu}
\affiliation{%
  \institution{SAMOVAR, T\'el\'ecom SudParis, Institut Polytechnique de Paris}
  \streetaddress{19 place Marguerite Perey}
  \city{Palaiseau}
%  \state{IdF}
  \country{France}
  \postcode{91120}
}

\renewcommand{\shortauthors}{Segovia, Rubio-Hernan, Cavalli, Garcia-Alfaro}
\renewcommand{\shorttitle}{A Survey on Cyber-Resilience Approaches for Cyber-Physical Systems}

\begin{abstract}
\CH{Concerns for the resilience of \gls*{cps} in critical infrastructure are growing. \gls*{cps} integrate sensing, computation, control, and networking into physical objects and mission-critical services, connecting traditional infrastructure to internet technologies. While this integration increases service efficiency, it has to face the possibility of new threats posed by the new functionalities. This leads to cyber-threats, such as denial-of-service, modification of data, information leakage, spreading of malware, and many others. Cyber-resilience refers to the ability of a \gls*{cps} to prepare, absorb, recover, and adapt to the adverse effects associated with cyber-threats, e.g., physical degradation of the \gls*{cps} performance resulting from a cyber-attack. Cyber-resilience aims at ensuring \gls*{cps} survival, by keeping the core functionalities of the \gls*{cps} in case of extreme events. The literature on cyber-resilience is rapidly increasing, leading to a broad variety of research works addressing this new topic. In this article, we create a systematization of knowledge about existing scientific efforts of making \gls*{cps} cyber-resilient.
We systematically survey recent literature addressing cyber-resilience with a focus on techniques that may be used on \gls*{cps}.  We first provide preliminaries and background on \gls*{cps} and threats, and subsequently survey state-of-the-art approaches that have been proposed by recent research work applicable to \gls*{cps}. In particular, we aim at differentiating research work from traditional risk management approaches, based on the general acceptance that it is unfeasible to prevent and mitigate all possible risks threatening a \gls*{cps}. We also discuss questions and research challenges,  with a focus on the practical aspects of cyber-resilience, such as the use of metrics and evaluation methods, as well as testing and validation environments.\\}
\end{abstract}




\maketitle

\section{Introduction}
\label{sec:intro}
Traditionally, the design of industrial systems was based on an isolation model, where the control of the operational technology was separated from the information technology. Today, both operational and information technology are integrated. Industrial physical processes are controlled by \CH{Cyber-Physical Systems (CPS)} that integrate modern computation and networking resources into traditional physical environments. They have emerged mainly in the industrial control system domain, using data acquisition and processing over networked control systems~\cite{ge_distributed_2017} to automate the remote execution of industrial tasks~\cite{zhang_survey_2016}.
Such integration has several advantages, for example, low maintenance costs, high reliability, flexibility, efficiency, and effectiveness to control the physical process~\cite{6305473}. The use of computation and networking resources to build a new generation of \gls*{cps} plays an important role in current critical nationwide infrastructures, such as electrical transmissions, energy distribution, manufacturing, supply chain, waste recycling, public transportation, health care, industrial process control, water infrastructure, and several others~\cite{ge_distributed_2017, lun_cyber-physical_2019}.

\CH{\gls*{cps} are composed of a physical process, sensors, actuators, and controllers. The sensors collect information about the physical process and send it to the controllers. Then, the controllers analyze the received information and calculate how to optimize the behavior of the physical process. As a result, the controllers send commands to the actuators to execute the corrective actions on the physical process. For example, to maintain the stability of the physical processes.} However, \gls*{cps} can be disrupted by cyber-physical attacks \cite{teixeira2012attack, Teixeira2015}, i.e., situations resulting from a cyber-attack, but manifesting physical effects, such as performance degradation~\cite{7954148}. These situations may put human safety at risk, cause harm in natural environments, interrupt industrial process continuity, and violate environmental regulations. Hence, cyber-physical attacks can lead to large economic losses, generate legal problems, and damage the reputation of the affected organizations~\cite{alguliyev_cyber-physical_2018}.
Many concerns have been raised about the vulnerabilities of control systems.
Recent history provides several cases of attacks on industrial infrastructures, which illustrate the threat that they represent. \CH{In particular, the security of industrial \gls*{cps} is drawing great attention after the Stuxnet malware~\cite{falliere2011w32,SPpanel2014} that considerably affected the performance of a uranium enrichment plant. The consequences of this event showed the dangers of successful cyber-threats carried out against \gls*{cps}.} Also, the well-known Ukraine attack \cite{case}  targeted power distribution networks causing outages as well as lasting damage. Another example is the Australian water services attacked by a disgruntled employee who infiltrated the system network and altered the control signals \cite{10.1007/978-0-387-75462-8_6}. The adversary took control of 150 sewage pumping stations resulting in the evacuation of one million liters of untreated sewage, over three months, into stormwater drains and onto local waterways. More examples of similar events can be found in \cite{sanchez_bibliographical_2019}.

Although pure cyber-attacks have shown limited damages to recent \gls*{cps}~\cite{HUANG200973}, full damages are feasible when considering adversaries that perpetrate control-theoretic manipulation, resulting from cyber-attacks, but leveraging physical disruption. This puts the focus on cyber-physical integrity attacks, which can rapidly move the system to unsafe states.
Ensuring the control of \gls*{cps} data exchanges is a challenging problem that requires a combination of both network and industrial control security. 
In addition, cyber-physical attacks may be hard to detect \cite{rubio2016nordsec,Rubio17ETT}. For this reason, resilience\footnote{In this article, we use the words \textit{resilience} and \textit{cyber-resilience} indifferently.} is especially relevant~\cite{10.1145/3462513}. Developing \gls*{cps} that can safely survive an attack is a current challenge~\cite{book_resilience_scott}.

Ensuring safety using only information security tools is not enough in the \gls*{cps} domain. Cybersecurity approaches do not cover all the possible vulnerabilities in the cyber components. For example, specific vulnerabilities may not have remediation mechanisms or they may be too expensive to implement. \CH{Even when the approach is implemented, detection algorithms are not free of false negatives and the remediation techniques may not be triggered.} As pointed out in \cite{8239925}, large research efforts have focused on intrusion detection for \gls*{cps}, but there is little discussion about what to do after the intrusion is detected, \ie, in remediation approaches that mitigate the effects of an attack. Most of the responses are manual or hardwired with a fixed response that cannot be configured. For this reason, attack tolerance should be enforced in critical systems to provide a correct service in the presence of successful attacks against the system \cite{RATHNAYAKA2022103123}. The resulting \gls*{cps} should satisfy high availability\footnote{In our work, availability means that legitimate users and processes have access to the system (and the resources of the system) whenever they need.} requirements to guarantee the execution of the critical tasks. It should be able to guarantee that the whole system remains operational even in the presence of attacks, even if that means working under graceful degradation modes. As a result, cybersecurity approaches should be complemented by secure control theory that provides attack models and a description of the interaction between the physical world and the control system. This will provide a better understanding of the attacks' consequences, development of new detection methods, response mechanisms, and architectures. It will also make the control systems more resilient to possible attacks and failures.

In this article, we focus on cyber-resilience techniques to build \gls*{cps} tolerant to cyber-physical attacks. We consider that the \gls*{cps} is a combination of cyber and physical components working together under discrete and continuous industrial environments \cite{ZHANG2013HighSystems}. We devote our work to protection techniques addressing networked control systems, \ie, a subset of \gls*{cps} dedicated to industrial control processes, usually performing critical functions. We analyze strategies that combine or have the potential to combine cybersecurity and control-theoretic approaches to build a solution that contemplates the cyber and the physical components of a \gls*{cps} to face the challenges created by cyber-physical adversaries. We differentiate research work from traditional risk management approaches, based on the general acceptance that it is unfeasible to prevent and mitigate all possible risks threatening a \gls*{cps}. We also discuss questions and research challenges,  with a focus on the practical aspects of cyber-resilience, such as the use of metrics and evaluation methods, as well as testing and validation environments.

\begin{figure*}[!h]
    \centering
    \includegraphics[width=.6\textwidth]{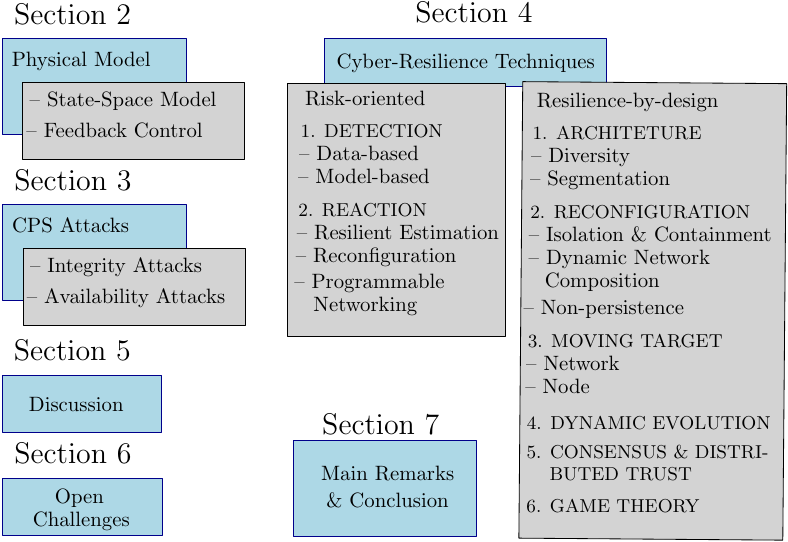}
    \caption{\CH{Organization of this article.}}
    \label{fig:classif}
\end{figure*}

\CH{The remainder of this article is outlined as follows.
Section~\ref{sec:ch2_CPS} explains how to model a \gls*{cps} and define the feedback control executed in the \gls*{cps} controller. Section~\ref{sec:CPS_attacks} provides a control-theoretic model of the cyber-physical integrity and availability attacks that we address in this article. Section~\ref{sec:ch2_techniques} provides our literature survey on cyber-resilience techniques to address the previously defined attacks. The selected literature was analyzed and classified based on risk-oriented techniques and resilience-by-design techniques. 
These two approaches are closely related. Remediation techniques are sometimes considered part of cyber-resilience. For this reason, we analyze the difference between them and we classify the collected proposals into two categories -- (1) \textit{detection and reaction} techniques, and (2) \textit{resilience-by-design} proposals. The approaches in each category are further classified into subcategories, as depicted in Fig.~\ref{fig:classif}.
Sections~\ref{sec:ch2_why_CT} and~\ref{sec:ch6_future_work} discuss research challenges and present open issues in the cyber-resilience area to lead future research. Section~\ref{sec:conclusion} concludes the article with our conclusions 
and main remarks.}

\section{Background on Cyber-Physical Systems}
\label{sec:ch2_CPS}

Cyber-Physical Systems (CPS), mathematically modeled in our work as networked control systems, are composed of distributed control systems and autonomous agents that need to make decisions in real time. They consist of two main parts. First, a cyber layer, containing the computing and network functionalities. Second, a physical layer, representing dynamic automation processes. Both together manage the distributed resources that monitor the behavior of physical phenomena and take the necessary actions to get control over them~\cite{ge_distributed_2017}. The \gls*{cps} becomes easier to automate at the cost of increasing the interaction between physical and cyber layers~\cite{zhang_survey_2016}. However, as a consequence, they become more vulnerable to new threats. Malicious actions in these systems are usually conducted by cross-layer adversaries that aim at harming the physical processes through the integration of physical and cyber layer attacks to cause, e.g., physical damages \cite{sanchez_bibliographical_2019}.

\gls*{cps} use a model able to manage and control the physical evolution of the system states. Controlling the states is a challenge since they follow the laws of the involved physical process, e.g., energy, water, or moving systems \cite{urbina2016survey}. For this reason, the physical properties of the system are used to create a model represented for the feedback control. This feedback control has to be able to regulate and manage the behavior of the system, i.e., a model able to confirm that the commands sent to the physical layer are executed correctly and the information coming from the physical states (through the sensors) is consistent with the predicted behavior of the system. 

In a \gls*{cps}, the {\it plant} (also referred to as \textit{system} by some authors) is the physical process that we want to control. The {\it actuators} perform physical actions over that process and the {\it sensors} collect the modifications produced at the physical layer. Using the data collected by the sensors, the feedback {\it controller} generates a residue between the data received from the sensors and the reference obtained after modeling the system. This residue, named {\it control error} by some authors, is used by the controller to create the {\it control input} to rectify, if necessary, the physical states using the actuators. The threat models explained in Section~\ref{sec:CPS_attacks} use some of the parameters and equations explained next.

\vspace{-.35cm}

\subsection{Physical Model}

How to obtain the model used in the feedback controller is a well-known problem in the control domain. Different techniques have been developed to provide a reference and generate the control input at each time step  \cite{ljung2010perspectives,error_estimated_TF_Goodwin,ljung1987system,ARX_ARMAX}; and also to create feedback control \cite{ricker1993model,barenthin2008complexity,Lee2004RobustEstimation}. The model can be obtained using a representation that relates to each possible input signal, the corresponding output signal. The two main mathematical approaches to model this are the \textit{transfer function} and the \textit{state-space model}. Both representations are equivalent since they are based on the differential equations that model the behavior of the physical process being controlled. 

Normally, a \gls*{cps} design process starts with the transfer function since it is the most direct form starting from the differential equations of the physical process. The transfer function $G(s)$ is the ratio of the Laplace transformation using the complex variable $s$ of the output $Y(s)$ to that of the input $U(s)$. It is represented as shown in Equation~\eqref{eq:trFn} by the division of two polynomials, the numerator is created by taking the coefficients $b_i$ of the output differential equation and the denominator using the coefficients $a_i$ of the input differential equation.

\vspace{-0.3cm}
\begin{equation}
  \label{eq:trFn}
G(s) =
    \dfrac
        {Y(s)}
        {U(s)} =
    \dfrac
        {\sum\limits_{i=0}^{m} b_is^{m-i}}
        {\sum\limits_{i=0}^{n} a_is^{n-i}}
\end{equation}

A transfer function with multiple inputs and multiple outputs is usually represented in a matrix which indicates the relationship of each input and each output of the system. Using well-known control theory techniques~\cite{Ogata}, it is possible to transform the transfer function into a state-space model by expressing the differential equations into matrices forms, cf. Equation~\eqref{eq:ch2_statespace} as follows:

\vspace{-0.6cm}
\begin{equation}
\centering
  \label{eq:ch2_statespace}
  \left.
  \begin{array}{ll}
       x_{k+1}=Ax_{k}+Bu_{k}+ w_{k} \\
       y_{k}=Cx_{k}+v_{k}
   \end{array}
   \right.
\end{equation}
where $x_{k}\in \mathbb{R}^n$ is the vector of the state variables at the $k$-th time step, $u_{k}\in \mathbb{R}^p$ is the control signal and $w_{k}\in\mathbb{R}^n$ is the process noise that is assumed to be a zero-mean Gaussian white noise with covariance $Q$, {\it i.e.} $w_k \sim N(0,Q)$. Controllers are normally implemented in a discrete form.

Moreover, $A\in \mathbb{R}^{n\times n}$ and $B\in \mathbb{R}^{n\times p}$ are respectively the {\it state} matrix and the {\it input} matrix.
The value of the output vector $y_{k} \in \mathbb{R}^m$ represents the measurements produced by the sensors that are affected by a noise $v_{k}$ assumed as a zero-mean Gaussian white noise with covariance $R$, {\it i.e.} $v_k \sim N(0,R)$ and $C\in \mathbb{R}^{m\times n}$ is the output matrix that maps the state $x_k$ to the system output.

%\vspace{-0.3cm}
\subsection{Feedback Control}

The previous equations define mathematically the behavior of a physical system. These equations are used by the feedback control to generate a closed-loop system. The output of the feedback control influences the input signal, e.g., to rectify the possible errors generated by the system. To build this type of feedback, two relevant mechanisms are \textit{Proportional-Integral-Derivative} (PID) controllers and \textit{Linear Quadratic Gaussian} (LQG) controllers.
LQG controllers provide feedback that holds better results than PID controllers \cite{LQG_PID}. LQG is a well-known technique for designing optimal dynamic feedback control laws. This optimal solution combines a Linear-Quadratic Estimator (LQE) with a Linear-Quadratic Regulator (LQR). These two components are independent but work together taking into account the measurement noise and process disturbance.
%(e.g., a Kalman filter)

The goal of an LQG controller is to produce a control law $u_k$ such that a quadratic cost $J$, that is a function of both the state $x_k$ and the control input $u_k$, is minimized:

\vspace{-0.3cm}
\begin{equation}
  J = \lim_{n \rightarrow \infty} E\left[\frac{1}{n}\sum_{i=0}^{n-1}(x_i^T \Gamma x_i + u_i^T \Omega u_i) \right]
  \label{eq:ch3_control_cost}
\end{equation}
where $\Gamma$ and $\Omega$ represent positive definite cost matrices~\cite{CDS_1998}.

It is well-known that a \textit{Kalman filter}-based LQE can be combined with a traditional LQR to solve the aforementioned control problem, as follows:

\begin{enumerate}
\item \textit{Kalman filter}-based LQEs use noisy measurements and produce an optimal state estimation $\hat x_k$ of $x$ (state);
\item the LQR, based on the state estimation $\hat x_k$, provides the control law $u_k$ that solves the problem (cf. Equation~(\ref{eq:ch3_control_cost})).
\end{enumerate}

A Kalman filter can estimate the state as follows:

\begin{itemize}
\item Predict (\textit{a priori}) system state $\hat{x}_{k|k-1}$ and covariance:

\vspace{-0.6cm}
  \begin{equation*}
    \hat{x}_{k|k-1}=A\hat{x}_{k-1} + Bu_{k-1}
  \end{equation*}
\vspace{-0.6cm}
  \begin{equation*}
    \label{eq:ch3_Covariance_error_apriori}
    P_{k|k-1}=AP_{k-1}A^T + Q
  \end{equation*}
\item Update parameters and (\textit{a posteriori}) system state and covariance:

\vspace{-0.6cm}
  \begin{equation*}
    \label{eq:ch3_Kalman_gain}
    K_{k}=(P_{k|k-1}C^T)(CP_{k|k-1}C^T + R)^{-1}
  \end{equation*}

 \vspace{-0.6cm}
  \begin{equation*}
    \hat{x}_{k}=\hat{x}_{k|k-1} + K_{k}(y_{k} - C\hat{x}_{k|k-1})
  \end{equation*}

 \vspace{-0.6cm}
  \begin{equation*}
    \label{eq:ch3_Covariance_error}
    P_{k}=(I - K_{k}C)P_{k|k-1}
  \end{equation*}
\end{itemize}
\noindent where $K_k$ and $P_{k}$ denote, respectively, the Kalman gain and the \textit{a posteriori} error covariance matrix, and $I$ is the identity matrix of appropriate dimensions.

The optimal control law $u_k$ provided by the LQR is a linear controller:
  $u_{k}=L\hat{x}_{k}$,
where $L$ denotes the feedback gain of the LQR that minimizes the control cost (cf.~ Equation~(\ref{eq:ch3_control_cost})), which is defined as follows \cite{Mo_2015}:

\vspace{-0.4cm}
\begin{equation*}
  L=-(B^{T}SB + \Omega)^{-1}B^{T}SA
\end{equation*}
with $S$ being the matrix that solves the following discrete-time
algebraic Riccati equation:

\vspace{-0.4cm}
\begin{equation*}
  S=A^{T}SA + \Gamma - A^TSB[B^{T}SB + \Omega]^{-1}B^{T}SA
\end{equation*}
%\vspace{-.8cm}

\section{Background on Cyber-Physical Threats}
\label{sec:CPS_attacks}

Control systems use safety mechanisms to handle failures and avoid accidents. Nevertheless, these control mechanisms cannot detect intentional malicious actions, such as cyber-physical attacks. Next, we present some existing cyber-physical adversary models and attack families.

\vspace{-0.2cm}
\subsection{Adversary Models}

The consequences of a successful cyber-physical attack can be more damaging than aggression on other networks because control systems are at the core of many critical infrastructures. We differentiate three main adversary models~\cite{rubio2017EurasipWatermak}:

\begin{itemize}
\item \textbf{Physical Adversary --} The adversary has physical access to the \gls*{cps} and can damage it by performing physical actions. For example, the adversary may cut the brakes of a connected autonomous car, destroy the valves that release the pressure in an industrial system, or perturb temperature sensor measurements by modifying their local surroundings \cite{weerakkody_resilient_2020, Teixeira2015}.
\item \textbf{Cyber Adversary --} The adversary can perform cybersecurity attacks (e.g., man-in-the-middle, buffer overflow, shell exploits, or others). The adversary has only knowledge about computation, storage, and network resources. Because of that, the attack can be easily detected by control-theoretic fault detection techniques~\cite{smith2015covert}.
Authors have systematized existing \gls*{cps} security research analyzing the taxonomy of threats, vulnerabilities, and attacks from the \gls*{cps} components perspective, with a special focus on cyber components~\cite{humayed_cyber-physical_2017} and cyber adversaries~\cite{alguliyev_cyber-physical_2018}. They also present the main difficulties and solutions in the estimation of the consequences of cyber-attacks, in terms of modeling, detection, and the development of security architectures.

\item \textbf{Cyber-Physical Adversary --} The adversary perpetrates cyber-attacks to cause tangible damage to physical components, for instance, by adding disturbances to a physical process via the exploitation of vulnerabilities in some computing and networking resources of the system.
The cyber-physical adversary is a combination of the two previous adversaries \cite{krotofil2015rocking}. First, the adversary uses a cyber-attack to gain position in the system from a remote location. Then, the adversary learns about the physical model to generate an attack with physical consequences but without being physically placed in the \gls*{cps} physical location. It can be hard to detect and locate a cyber-physical adversary, whose attacks may often be confused with faults in the system.
\end{itemize}

\subsection{Attack Families}
\label{sec:ch2_attack_taxonomy}

Different cyber-physical attack families have been reported in the literature. Authors in \cite{HUANG200973} provide control-theoretic models for integrity and denial-of-service (DoS) attacks. Similar techniques have been reported in  \cite{DIBAJI2019394}, naming them deception and disruption cyber-physical attacks, respectively. The work in \cite{HUANG200973} shows that a traditional DoS attack does not have a significant effect when the system is in a steady state. However, the violation of integrity properties in such attacks can rapidly move the system to unsafe states.

A convenient attack classification in the existing literature is the one proposed in \cite{Teixeira2015}, which introduced the attack space as a three-dimensional graphical characterization of the attacks. It considers the following three dimensions: the adversary’s a priori knowledge of the system’s model, the disruption of resources, and the disclosure of resources. The knowledge of the system’s model allows the adversary to develop sophisticated attacks, which have more severe consequences and are harder to detect with traditional approaches. The disclosure of resources lets the adversary to obtain sensitive information, which may be used to generate knowledge about the system, but cannot be used to disrupt the system's operation. Finally, the disruption of resources can be used to affect the system's operation (e.g., maintaining the stability of the system).

\begin{figure}[!hptb]
\begin{center}
\includegraphics[width=\columnwidth]{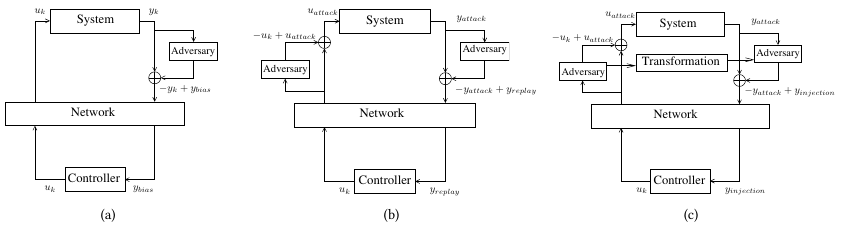}
%\vspace{-0.6cm}
\caption{(a) Stealth attack. (b) Replay attack. (c) Covert attack. \label{fig:ch2_all}}
\end{center}
\vspace{-0.6cm}
\end{figure}

Fig. \ref{fig:ch2_all} depicts block diagrams representation of cyber-physical adversaries attacking a control loop. The $\bigoplus$ symbol represents a \textit{summing junction}, \ie, the sum of input signals.
To take control of the physical process, the adversary may send a malicious command $u_{attack}$ to the \textit{System} that will be executed by the actuators. After that, to deceive the controller and go unnoticed, the adversary may modify the sensors' readings $y_{attack}$ to inject a measurement value $y$. The adversary may use a combination of different commands $u$ and measurements $y$ to deceive the controller and damage the system.

Next, we outline some cyber-physical attacks following the taxonomy presented in \cite{Teixeira2015}. Cyber-physical adversaries use integrity attacks to exploit vulnerabilities in the control mechanism and take control of the physical process. For this reason, all the attacks are assumed to inject malicious traffic. However, they are classified into different categories because they exploit different vulnerabilities in the control loop. As a consequence, these attacks produce different effects on the physical process and they may require different approaches to be solved.

\subsubsection{False-Data Injection Attack or Stealth Attack}

In this attack family (cf. Fig.~\ref{fig:ch2_all}(a)), the adversary modifies some sensor readings by applying physical interference, at the sensor device, or by perturbing the communication channel to disrupt the system \cite{Cardenas_Risk_Detection, teixeira2012attack, sanchez_bibliographical_2019}.
To carry out attacks from this family, the adversary needs knowledge about the behavior of the system, such as the system dynamics, the command signals, and the control detection threshold. The adversary drives slowly the control decisions out of the correct behavior and produces wrong control decisions to cause a malfunction in the system. From a control-theoretic perspective, the injected false data should not affect the system residues (cf. Section~\ref{sec:ch2_CPS}). This means that the injected data should not alter the sensor measurement variations. Otherwise, the attack would be easily detected.

\subsubsection{Replay Attack}

Fig.~\ref{fig:ch2_all}(b) shows adversaries conducting a cyber-physical replay attack by modifying some sensor readings (e.g., by replicating previous measurements, corresponding to normal operating conditions). Then, the adversaries modify the control input to affect the system state. These adversaries are not required to know the system process model, but  access to all the sensors is required to carry out a successful attack. This type of adversary is undetectable with a monitor detector which only verifies sensors' measurements. To detect the attack, it is required to add some protection to the input control signal $u_k$ \cite{Mo_2014}, defined in Section~\ref{sec:ch2_CPS}.

\subsubsection{Covert Attack}

Adversaries, depicted in Fig.~\ref{fig:ch2_all}(c), read and add to both, the control data and the sensor measurements. The difference with the replay attack is that the adversary needs \textit{a priori} knowledge about the system process to create a transformation that is correlated with the control model, i.e., the attack requires knowing the behavior of the physical system as well as the behavior of the feedback control. This type of adversary is considered undetectable if measurements are compatible with the physical process. In other words, the attack cannot be distinguished from the regular system operations \cite{smith2015covert}.

\subsubsection{DoS Attack}

A denial-of-service (DoS\footnote{We can also consider distributed denial-of-service (DDoS), where multiple nodes attack one or many other components.}) aims at disrupting the communication between the remote elements (e.g., elements related to supervisory control and data acquisition protocols) and local elements closely related to the system (e.g., terminal units and  programmable logic controllers connected to the sensors and actuators of the system), hence disrupting the availability of feedback control~\cite{Dos_Yuan}. By disconnecting the controller from the physical device, it is possible to avoid the process monitoring and make the system vulnerable to other malicious actions \cite{Dos_injectionAt_Wei}.

It is worth noting that cyber-physical DoS attacks are launched using integrity attacks to cause significant damage. In this case, the attack compromises the integrity of the messages, as shown in Fig.~\ref{fig:ch2_all}(b), with two objectives. First, to disrupt the communication between the controller and the system, generating a loss of system's supervision that may be not easy to detect. Second, to inject malicious messages to move the system from the stability point. This way, the adversary generates an unavailability of the system to the authorized users to make it available just for the malicious actions. As a result, this adversary affects the integrity of the system to generate also an availability problem.

\subsubsection{Command Injection Attack} This attack uses the protocols and device vulnerabilities to inject false commands into the control systems to disrupt control actions or system settings. It is similar to the attack shown in Fig.~\ref{fig:ch2_all}(a), but the adversary injects the malicious traffic in the control command, i.e., in $u_k$. For example, by overwriting the remote registers associated with some supervisory control or exploiting the data acquisition protocols  \cite{gao2014cyber}.

\subsubsection{Zero Dynamics Attack}

This attack family assumes vulnerabilities present in the dynamics of the system concerning properties used to monitor and control the behavior. This attack is similar to the command injection attack, but it makes an unobservable state unstable and disrupts this unobservable part of the system without being detected by the controller~\cite{Teixeira2015, Dynamic_attackChen}. A solution to avoid this kind of attack is to update the architecture of the system to make all the states observable, e.g., by deploying more sensors to avoid unobservable situations in the system.

As we have seen in this section, the existence of availability and integrity vulnerabilities is the main security issue in \gls*{cps}. Although pure cyber-attacks may have a limited impact on the system, combined with control-theoretic strategies may cause important physical damages~\cite{HUANG200973}. Indeed, cyber-physical integrity attacks can rapidly move the system to unsafe states. Also, cyber-physical DoS attacks can benefit from integrity issues, to cause significant damages. In this case, the integrity of the messages is compromised with two objectives. First, to disrupt the communication between the controller and the system, hence leading to supervision loss (which is hard to detect). Second, to inject malicious messages to move the system from its stability point. This way, the adversary generates unavailability of the system to authorized users, e.g., to make it available just for the malicious actions. In the next section, we present existing resilience techniques to face cyber-physical adversaries and reduce the impact they may have on the system safety.

\section{Systematic Survey on Cyber-Resilience Literature}
\label{sec:ch2_techniques}

Cyber-resilience is the ability of a system to \emph{prepare, absorb, recover}, and \emph{adapt} to adverse effects \cite{book_resilience}.
The \emph{preparation} phase is characterized by identifying the critical functions or services and stakeholders.
It is important to understand the critical functionalities to guide the planning actions. The \emph{absorption} phase involves the capacity of the system to contain the attack under degraded performance. It is the ability of a system to tolerate the stress. Thresholds are important to determine whether a system can absorb a shock or not. During the \emph{recovery} phase, the system starts the process to restore its normal behavior as quickly and efficiently as possible. Finally, the \emph{adaptation} phase involves a postmortem evaluation to improve the response and learn from past experiences.

Although the previously mentioned definition provides a clear view of the resilience stages, it may also be too broad for the \gls*{cps} domain.
A given \gls*{cps} with unlimited resources (e.g., unlimited time) will eventually recover from all failures and attacks. Hence, resilience should be established considering a minimum group of conditions, e.g., in terms of temporal and computational resources. Under this assumption, and with the \gls*{cps} context in mind, a more appropriate definition of resilience points out 
the necessity of providing~\cite{clark_cyber-physical_2019}: (1) full correctness maintenance of the core set of crucial functionalities despite ongoing adversarial misbehavior (i.e., it is acceptable for non-crucial functionalities to be affected temporarily, such as partially degraded or complete failure); and (2) guaranteed recovery of the normal operation of the affected functionalities within a predefined cost limit. In addition, attack tolerance and graceful degradation are two properties that we may want to satisfy in a resilient system. Attack tolerance assumes that attacks can happen and be successful. The overall system must remain operational and provide a correct service. Graceful degradation is the ability of a system to continue functioning even in a lower performance after parts of the system have been damaged, compromised, or destroyed. The efficiency of the system working in graceful degradation usually is lower than the normal performance. It may decrease as the number of failing components grows. The purpose is to prevent a catastrophic failure of the system.

%%%%%%%%%%%%%%%%%%%%%%%%%%%%%%%%%%%%%%%%%%%%%%%%%%%%%%%%%%%%%%%%%%
%%%%%%%%%%%%%%%%%%%%%%%%%%%%%%%%%%%%%%%%%%%%%%%%%%%%%%%%%%%%%%%%%%

\begin{table}[!b]
\vspace{-0.4cm}
\begin{center}
\caption[Resilience Approaches for CPS]{\label{tab:techniques}Proposed resilience approaches for CPS. (Top) Resilient Control Techniques. (Down) Cyber-Resilience Techniques.}
\small
%%%%%%%%%%%%%%%%%%%%%%%%%%%%%%%%%%%%%%%%%%%%%%%%%%%%%%%%%%%%%%%%%%
\begin{tabular}{| p{5.6cm} | c |c  |c |c |c | c |p{4.1cm} |}\hline
 \multirow{2}{*}{\textbf{\Longunderstack{Resilient Control Techniques\\Section \ref{resControl}}}} &
    \multicolumn{4}{c|}{\textbf{Layer}} & \multirow{2}{*}{\textbf{Proposals}} \\ \cline{2-5}
    &

    \rotatebox{90}{\textbf{Physical~}} &
    \rotatebox{90}{\textbf{Network~}} &
    \rotatebox{90}{\textbf{Control}} &
    \rotatebox{90}{\textbf{Cyber}} & \\ \hline\hline

%Redundancy & &\checkmark & &\checkmark &\checkmark & & & \cite{ashraf_capacity-aware_2018} \\ \hline
\textbf{Detection} & \multicolumn{4}{c|}{}&\\ \hline

        Data-based Approach & &\checkmark  & & \checkmark & \Longunderstack{\cite{10.5555/1162264}, \cite{shawe-taylor_cristianini_2004}, \cite{Hofmann_ML}, \cite{10.1145/2542049}, \cite{cheminod_review_2013}, \cite{6942184}, \cite{ahmed_survey_2015}, \cite{ding_survey_2018}, \cite{6786081}}\\ \hline

        Model-based Approach & &  &\checkmark & & \Longunderstack{\cite{Mo_2015}, \cite{Miao2013StochasticDetection}, \cite{rubio2017EurasipWatermak}, \cite{do2014statistical}, \cite{arvani2014detection}, \cite{Correlation_detectorLokhov}, \cite{Wang2014},\cite{anomaly_detectionChen}, \\  \cite{6307833}, \cite{detection_using_modelbased_Dehghani},  \cite{Zhu2015Game-theoreticSystems}, \cite{bobbadetecting}, \cite{pasqualetti2015control}, \cite{luenberger_introduction_1971}, \cite{shoukry_event-triggered_2016}, \cite{schellenberger_detection_2017}, \cite{weerakkody_resilient_2020}}\\ \hline \hline

\textbf{Reaction} & \multicolumn{4}{c|}{}& \\ \hline \hline

    Resilient State Estimation & &  & \checkmark&  & \Longunderstack{\cite{weerakkody_resilient_2020}, \cite{fawzi_secure_2014}, \cite{pajic_robustness_2014}, \cite{pajic_attack-resilient_2017}, \cite{6881627}, \cite{doi:10.1080/00207721.2014.906683}, \cite{weimer_attack-resilient_2014}, \cite{shoukry_secure_2017}, \cite{mishra_secure_2017}, \cite{10.1145/1995376.1995394}, \\ \cite{7299903}, \cite{6730927}, \cite{TAN2017313}}\\ \hline

    Reconfiguration & \checkmark& \checkmark & \checkmark& \checkmark & \Longunderstack{\cite{giraldo_security_2017}, \cite{8443136}, \cite{Dos_Yuan}, \cite{reflect_Ana}, \cite{8530771}, \cite{10.1145/3232848}, \cite{6425820}, \cite{DIBAJI2019394}, \cite{Cetinkaya2016}, \\ \cite{YANG2017145}, \cite{LEI2016286} } \CH{\cite{Sun__event_triggered_2022}}\\ \hline

    Programmable Networking & &\checkmark  & &  & \Longunderstack{\cite{Campbell_opensignal}, \cite{Tennenhouse_active_network}, \cite{Netconf}, \cite{Kreutz_SDN_survey}, \cite{sahay}, \cite{hadega}, \cite{itl2018}, \cite{molina_software-defined}, \cite{PIEDRAHITA2018}}\\
    \hline
    \hline
\end{tabular}

\vspace{0.15cm}
%===============================================================

\begin{tabular}{| p{4cm} |  c | c | c || c |c  |c |c |c | c |p{3.8cm} |}\hline
 \hline
 \multirow{2}{*}{\textbf{\Longunderstack{Cyber-Resilience Techniques\\Section \ref{subsec:ch2_resilience_sota}}}} &
    \multicolumn{3}{c|}{\textbf{Phase}} &
    \multicolumn{4}{c|}{\textbf{Layer}} & \multirow{2}{*}{\textbf{Proposals}} \\ \cline{2-8}
    &
    \rotatebox{90}{\textbf{Absorb}} &
    \rotatebox{90}{\textbf{Survive}} &
    \rotatebox{90}{\textbf{Recover}} &
    \rotatebox{90}{\textbf{Physical~}} &
    \rotatebox{90}{\textbf{Network~}} &
    \rotatebox{90}{\textbf{Control}} &
    \rotatebox{90}{\textbf{Cyber}} & \\ \hline\hline

\textbf{Architecture Design} & \multicolumn{7}{c|}{}&\\ \hline

        Diversity & &\checkmark & &\checkmark & \checkmark & & \checkmark & \Longunderstack{\cite{larsen_sok_2014}, \cite{8029792}, \cite{532621}, \cite{chaves_improving_2017}, \cite{COHEN1993565}, \cite{Forrest_buildingdiverse}, \cite{10.1145/2508859.2516675},\\ \cite{Jackson2011}, \cite{10.1007/978-3-642-00730-9_10}, \cite{10.1007/978-1-4614-5416-8_8}, \cite{10.1007/978-3-540-70542-0_1}, \cite{10.1145/948109.948146}, \cite{6494997}, \cite{cispa450}}\\ \hline

        Segmentation &\checkmark & & & & \checkmark & & & \cite{gengeSegmentation}, \cite{10.1007/978-3-642-41488-6_12}\\ \hline \hline

\textbf{Reconfiguration} & \multicolumn{7}{c|}{}&\\ \hline

    Isolation and Containment &\checkmark & & &\checkmark & \checkmark & &\checkmark & \Longunderstack{\cite{bellini_cyber_2019}, \cite{chen_robustness_2020}, \cite{avizienis_concept_2016}, \cite{haque_modeling_2019}, \cite{kwasinski_modeling_2020}, \cite{xu_islanding_2020}}\\ \hline

    Dynamic Network Composition & \checkmark & \checkmark & & & & & & \Longunderstack{\cite{PIEDRAHITA2018}, \cite{segovia_reflective_2020}, \cite{januario_distributed_2019},  \cite{marshall_context-driven_2019}, \cite{chen_adaptive_2020}} \\ \hline

    Non-Persistence & \checkmark & \checkmark & & & & & & \cite{9147232}, \cite{PRADHAN2016344} \\ \hline \hline

\textbf{Moving Target Defense (MTD)} & \multicolumn{7}{c|}{}&\\ \hline

Network MTD & \checkmark & \checkmark & \checkmark & & \checkmark &  &  & \Longunderstack{\cite{kanellopoulos_moving_2019}, \cite{10.1007}, \cite{ANTONATOS20073471}, \cite{8390877}, \cite{10.1145/2663474.2663479}, \cite{Macfarl_thesdn},  \cite{6924217}, \cite{Aseeri}, \\ \cite{6682715},  \cite{978-3-319-50011}, \cite{lei_moving_2018}, \cite{zheng_survey_2019}, \CH{\cite{Bradley_Potteiger_2022},
\cite{Xiaoyu_moving_2022}, \cite{Azab_moving_2022}}}\\ \hline

Node MTD & \checkmark & \checkmark & \checkmark & & & \checkmark & \checkmark & \Longunderstack{\cite{kanellopoulos_moving_2019}, \cite{griffioen_moving_2019},\cite{giraldo_moving_2019}, \cite{10.1145/2663474.2663479}, \cite{lei_moving_2018}, \cite{zheng_survey_2019}, \cite{9266030}, \cite{giraldo_moving_2019}, \\ \cite{weerakkody_moving_2016},  \cite{9266030} \CH{\cite{giraldo_moving_2022}, \cite{Liu_moving_2021}
}}\\ \hline \hline

\textbf{Dynamic Software Evolution} & \checkmark & \checkmark & & & & & \checkmark &  \cite{10.1016/j.cose.2011.08.007}, \cite{reflect_Ana}, \cite{he_software_2008}, \cite{kon_case_2002}\\ \hline \hline

\textbf{Consensus \& Distributed Trust} &\checkmark & \checkmark & & & \checkmark &\checkmark & & \Longunderstack{\cite{5605238}, \cite{severson_resilient_2020}, \cite{wen_distributed_2018}, \cite{mahmoud_distributed_2013}, \cite{amini_performance_2019}, \\ \cite{saldana_resilient_2017}, \cite{meng_studies_2014}, \cite{yan_resilient_2020}, \cite{usevitch_resilient_2019}, \cite{shabbir_resilient_2020}, \cite{zegers_event-triggered_2019}} \\ \hline \hline

\textbf{Game Theory} & \checkmark & \checkmark & & & & & \checkmark & \Longunderstack{\cite{hasan_game-theoretic_2020}, \cite{huang_dynamic_2020},
\cite{sanjab_bounded_2016}, \cite{kanellopoulos_non-equilibrium_2019}, \cite{zhu_game-theoretic_2013}, \cite{rao_resilience_2015}}\\ \hline \hline
\end{tabular}
\end{center}
%\vspace{-0.8cm}
\end{table}

\subsection{Risk Management vs. Cyber-Resilience}
\label{sec:risk_resilience}

Risk management and resilience are different but related concepts~\cite{arghandeh_definition_2016}. Although they are both grounded in a similar mindset (e.g., reviewing systems for weaknesses and identifying policies or actions that could mitigate or resolve such weaknesses), substantial differences exist \cite{linkov_science_2019}.

On the one hand, a risk is assessed by the likelihood of an undesirable event and the consequence of that event using probability distribution functions. On the other hand, resilience is about the remediation of unexpected rare extreme failures, whose likelihood cannot be estimated from historical data. Risk management is concerned with analyzing threat-by-threat to derive a precise quantitative understanding of how a given threat generates harmful consequences. Such exercise works well when the threats are categorized and understood, yet develops limitations when working with complex interconnected systems. Building from this limitation, resilience complements traditional risk-management approaches by reviewing how systems perform and function in a variety of scenarios, agnostic of any specific threat.

In addition, resilience requires thinking in terms of how to manage systemic, cascading effects on other directly and indirectly connected nodes. While risk management centers around the probability of hitting the weak points of a system, resilience is grounded in ensuring system survival. It finds strategies to keep the functionality of the core system in the face of extreme events. Hence, resilience is based on a general acceptance that it is virtually impossible to prevent or remediate all categories of risk simultaneously, and before they occur~\cite{book_flammini}.

New adversary models, as those presented in Section \ref{sec:CPS_attacks}, create new challenges to achieve
resilient systems. Indeed, achieving security in a \gls*{cps} requires solutions that extend beyond what is offered by state-of-the-art cybersecurity products. As a result, a new research area must focus on strategies to face cyber-physical adversaries. In the control-theoretic community, this new area is known as \textit{Resilient Control}~\cite{weerakkody_resilient_2020}. It is worth noting that although these approaches are called \textit{resilient} by control theorists, from a cybersecurity standpoint, resilient control is still dependent on a \emph{detection} and \emph{reaction} paradigm. In other words, although resilient control incorporates in the traditional fault-tolerant control new strategies to face cybersecurity breaches, it still aims at determining how a controller can detect, correctly estimate the system state, and recalculate the required commands despite malicious data. It also aims at responding to the attacks with appropriate countermeasures, to achieve stability and graceful degradation while the system is under attack. This objective can be achieved through a system theoretical analysis of the \gls*{cps}.

Next, we present our survey of solutions in both categories. First, we survey in Section~\ref{resControl} \emph{resilient control} techniques under the traditional \emph{detection} and \emph{reaction} paradigm.
Then, we survey in Section~\ref{subsec:ch2_resilience_sota} \emph{cyber-resilience} techniques that provide system recovery without triggering any additional behavior. Our literature survey is summarized in Table~\ref{tab:techniques}.

\vspace{-.25cm}

%======================================
%%      DETECTION
%======================================
\subsection{Resilient Control}
\label{resControl}

Detection and mitigation for cyber-physical attacks are not trivial. It requires incorporating control-theoretic strategies into traditional cybersecurity approaches to contemplate the new vulnerabilities. In this section, we present resilient control strategies based on detection and reaction mechanisms for \gls*{cps}.

\subsubsection{\textbf{Detection Approaches}}
\label{detection}

There are two main strategies for attack detection in \gls*{cps}: \textit{data-based} and \textit{model-based} approaches~\cite{7954148}. Data-based and model-based approaches are complementary solutions, together they consider the interaction between both cyber and physical layers.

\subsubsubsection{Data-based Approach}

This approach does not require system and attack models for detection. It is based on traditional machine learning and pattern recognition techniques \cite{10.5555/1162264, shawe-taylor_cristianini_2004, Hofmann_ML} for analyzing hidden patterns in the observed training dataset, for example, control signals and sensor measurements. Mitchell \etal \cite{10.1145/2542049}, Cheminod \etal \cite{cheminod_review_2013}, and Han \etal \cite{6942184} provide surveys of intrusion detection techniques focusing only on data-based approaches using traditional intrusion detection systems. Ahmed \etal \cite{ahmed_survey_2015} provide a survey of trust-based detection and isolation approaches for malicious nodes in sensor networks. In addition,  Ding \etal \cite{ding_survey_2018} survey the development of attack detection for industrial \gls*{cps} and discuss control and state estimation in the case of an attack. Also, Beaver \etal \cite{6786081} provide an evaluation of machine learning methods to detect malicious communications in supervisory control and data acquisition protocols.

\textit{Advantages and limitations:} This detection technique considers cyber and network patterns to identify attacks. For this reason, it can detect cyber-attacks. However, it is not able to detect all kinds of cyber-physical attacks, since it does not consider the control model. It has a partial view that does not include the physical components.

\subsubsubsection{Model-based Approach} This approach uses the model of the systems to detect attacks. The decision is based on the comparison between system observations and model outputs. The system is under attack if the observed data is no longer consistent with the estimated outputs of the normal mode. This comparison may not be obvious because of the presence of model uncertainties, nuisance parameters, and random noise.

There are five main strategies for control-theoretic model-based attack detection \cite{rubiohernan:tel-01810321}: \textit{watermark-based detectors}, \textit{signal-based detectors}, \textit{state relation-based detectors}, \textit{cross layer-based resilient detectors}, and \textit{auxiliary systems detectors}. Next, we summarize the main ideas underlying each strategy.

\begin{itemize}
\item
% Watermark-based detector
In the case of \textit{watermark-based detectors}, a low amplitude noise, called watermark, is added to the control measurements to verify, by using a detection mechanism, that the sensor measurements and commands are not modified, i.e., the control measurements with the watermark have to be correlated with the sensor measurements. For example, Mo \etal \cite{Mo_2015} propose the use of Kalman filters to detect cyber-physical replay attacks by adapting traditional failure detection mechanisms via watermarking. Miao \etal \cite{Miao2013StochasticDetection} improve the performance of the aforementioned detection mechanism using a stochastic game approach. The work has also been improved by Rubio-Hernan \etal \cite{rubio2017EurasipWatermak} to incorporate more advanced adversaries capable of learning the physical model. In the same way, Do \etal \cite{do2014statistical} propose a detection approach based on the knowledge of the system's behavior and its stochastic variations to detect data manipulation.

\item
% Signal-based detector
\textit{Signal-based detectors} use the signal statistical properties and the system behavior to detect attacks. For example, Arvani \etal \cite{arvani2014detection} describe a model to detect and identify random signal data-injections attacks. It is based on discrete wavelet transform analysis to exploit the statistical properties of the signal and the dynamic model of the system. It also uses a chi-square detector to identify anomalies. Lokhov \etal \cite{Correlation_detectorLokhov} present a protocol for detection and localization of disturbance based on a special correlation matrix. The matrix allows: (1) detecting anomalies using spectral methods; (2) localizing a subset of anomalous nodes within the system; and (3) identifying the functional role of the inferred anomaly based on the sensor labels.

\item
% State relation-based detector
\textit{State relation-based detectors} use the correlation of system states and the system behavior, to identify anomalies. For example, Wang \etal \cite{Wang2014} propose a relation-graph-based detector scheme to detect false data injection attacks, even when the injected data may seemly fall within a valid and normal range. A correlation model extracts the relation among the different variables of the system to create a graph model with the possible valid system states. The correlation model uses a forward correlation that is not affected by time and a feedback correlation that depends on time. Chen \etal \cite{anomaly_detectionChen} present a distributed anomaly detection algorithm using graph theory and spatiotemporal correlations to analyze the physical process in real-time. Amin \etal \cite{6307833} develop a model-based scheme for detection and isolation. The scheme is based on a group of unknown input observers designed for a linear delay-differential system obtained as an analytically approximate model. The generated conditions are delay-dependent, and can also incorporate communication network-induced time-delays in the sensor-control data. To detect and isolate the attacks, they use a residual generation procedure. Also, Dehghani \etal \cite{detection_using_modelbased_Dehghani} present a static state estimation algorithm able to detect  integrity attacks against smart grids.

\item
% Cross layer-based resilient detectors.
\textit{Cross-layer based resilient detectors} combine control and cyber techniques in a single intrusion detection system. For example, Zhu \etal \cite{Zhu2015Game-theoreticSystems} propose a game-theoretic framework that integrates the discrete-time Markov model for modeling the evolution of cyber states with continuous-time dynamics describing the controlled physical process. The cross-layer design is created between physical and cyber detection layers to maximize the chances of identifying security events. Bobba \etal \cite{bobbadetecting} show that protecting only a set of basic measurements is enough to detect attacks against physical and network malicious actions. In addition, Pasqualetti \etal \cite{pasqualetti2015control} use geometric control theory to optimize cross-layer resilient control systems. They conclude that by using a geometric model of the system, it is possible to detect and estimate the system state in the presence of unknown inputs.

\item
%Luenberger observers-based detectors
\textit{Auxiliary system detectors} use state observer techniques (e.g., Luenberger observers \cite{luenberger_introduction_1971}) to build a digital copy of the system and be able to control its behavior. For example, Shoukry and Tabuada~\cite{shoukry_event-triggered_2016} describe an algorithm for state reconstruction from sensor measurements that are corrupted using a Luenberger observer.  Also, Schellenberger \etal~\cite{schellenberger_detection_2017} extend an original plant with an auxiliary system that does not add additional delay into the system. The auxiliary system is designed as a linear discrete-time digital copy with similar dynamics to the original system, but capable of conducting attack detection. For this detection strategy, a model of the overall system dynamics and the switching signal of the auxiliary system are needed. The residuals of the Luenberger observer are then monitored for deviations from zero, which indicates the existence of attacks.
\end{itemize}

\textit{Advantages and limitations:} This detection technique considers the physical model to identify attacks. It is suitable to identify cyber-physical attacks using feedback control. However, the information on traffic patterns and cyber-attacks identification may be limited. For this reason, it is complementary to the previous approach which is based on cyber and network data. Both techniques working together, have a more complete and integral view of the system.

%======================================
%%      REACTION
%======================================

\subsubsection{\textbf{Reaction Approaches}}
\label{reaction}

As pointed out in \cite{8239925}, large research efforts have focused on intrusion detection. There is little less discussion about what to do after the intrusion is detected, \ie, in remediation approaches that mitigate the effects of an attack. Most of the responses in \gls*{cps} are manual or hardwired with a fixed response that cannot be configured. In the sequel, we survey some representative proposals under the reaction (after detection) paradigm.

% Resilient state estimation
\subsubsubsection{Resilient State Estimation} When an adversary modifies data, system recovery requires knowing the real state of the system. For this reason, resilient state estimation is a technique that can help in terms of system reaction. It allows a remote defender to maintain an understanding of the system state under attack, even when a subset of inputs and outputs are under the control of an adversary \cite{weerakkody_resilient_2020}. As a result, the defender can still have reliable state information to apply an appropriate feedback control law, to better understand the portions of the system that have been compromised, and to design attack-specific countermeasures.

\CH{Approaches for resilient state estimation can be found in the following literature. Fawzi \etal \cite{fawzi_secure_2014} propose an efficient state reconstructor inspired by techniques used in compressed sensing and error correction over real numbers. They also characterize the maximum number of attacks that can be detected and corrected as a function of the system state matrices. Pajic \etal \cite{pajic_robustness_2014} present a method for state estimation in the presence of attacks, for systems with noise and modeling errors such as jitter, latency, and synchronization problems that are mapped into parameters of the state estimation procedure. Pajic \etal \cite{pajic_attack-resilient_2017} also propose a state estimation approach in the presence of bounded-size noise for sensor attacks where any signal can be injected via compromised sensors.} 

\CH{In addition, Mo and Sinopoli \cite{6881627} propose a state estimator based on $m$ measurements that can be potentially manipulated by an adversary. The adversary is assumed to have full knowledge about the true value of the state to be estimated and about the value of all the measurements. If the adversary can manipulate up to $l$ of the $m$ measurements, then the estimator works properly when the adversary compromises less than half of the measurements, \ie, $(l < m/2)$. The solution is formulated as an optimization problem where one seeks to construct an optimal estimator that minimizes the worst-case expected cost against all possible manipulations by the adversary. Keller \etal \cite{doi:10.1080/00207721.2014.906683} propose a state estimation of stochastic discrete-time linear systems in the case of malicious disturbance that switches between unknown input and constant bias. This means that when corrupted control signals are received by the controller, detectors based on Kalman Filters are used to estimate the state of the system and the exogenous unknown input of the system (i.e., the malicious inputs). In addition, the malicious control signal is blocked at the occurrence of data losses, and the unknown input is transformed to a constant bias at the input of the system. Weimer \etal \cite{weimer_attack-resilient_2014} introduce a resilient estimator for stochastic systems using a mean squared error for the state that remains finitely bounded and is independent of attacks in measurements.} 

Shoukry \etal \cite{shoukry_secure_2017} and Mishra \etal \cite{mishra_secure_2017} propose secure state estimation algorithms for linear dynamical systems under sensor attacks and in the presence of noise. The approaches are based on satisfiability modulo theory, which is a technique used to express problems that should satisfy constraints, \ie, decision problems using logical formulas expressed in first-order logic \cite{10.1145/1995376.1995394, 7299903}.
Another technique used to improve the state estimation accuracy is to consider multiple sensor systems instead of one single sensor system \cite{6730927, TAN2017313}. In this case, data fusion is a process in which the received data is integrated from different sensors observing the same system.

\textit{Advantages and limitations:} This approach is useful when sensors, actuators, or network traffic have been compromised. It provides a reliable state of the system even when an adversary injects malicious traffic into the system. As a result, this technique helps the system recovery because it allows maintaining an understanding of the state under attack, even when a subset of inputs and outputs are malicious. The limitation is that it can only repair a maximum number of compromised values. In addition, it is hard to ensure that the control commands are executed correctly by simply using state estimation techniques. For that, complementary actions need to be included in the response plan and to ensure that the estimated data properly reaches its destination.

\vspace{-.25cm}

\subsubsubsection{Reconfiguration} Once the system is compromised, it is required to ensure that the control commands arrive correctly to the actuators. One possibility to do this is to alter dynamically the configuration of the system to minimize the effects of the attack. For example, changing the network topology, configuration of the devices, firewall rules, or quarantining (rerouting) traffic. In other words,  the system structure is modified to face the attacks. For instance, one option would be to increase the number of sensors such that attacks are identified faster or add extra layers of security to those elements that are more vulnerable to cyber-attacks \cite{giraldo_security_2017}. Components may also be isolated. Li \etal \cite{8443136} propose a decision-making approach for intrusion response aiming to determine the optimal security strategy against the attacks. The strategy tries to secure attack paths with higher priority, in addition to responding to functional failures. Authors assess both cyber and physical domains with an in-depth analysis of attack propagation. Yuan \etal \cite{Dos_Yuan} propose a resilient controller design for \gls*{cps} under DoS attacks. The proposal uses a framework that incorporates an IDS and robust control. The robust control in the physical layer is based on an algorithm with value iteration methods and linear matrix inequalities, e.g., for computing the optimal security policy and control laws. The cyber state is modeled as a continuous Markov process to defend against malicious behavior.

Other techniques incorporate dynamically new on-demand capabilities  to face the attacks. For example, using pre-configured virtual machines to help affected components, adding new cloud-based services to help with denial-of-service attacks, or distributing tasks in a different organization.
Ismail \etal \cite{8530771} propose an optimization of the defense countermeasures deployment. To design the approach, the available information is presented in an attack graph, representing the evolution of the state of the attacker in the system. Then, they find the optimal security policy to  maximize the system protection using Markov decision processes. This way, countermeasures are prioritized to respond efficiently to the intrusion. Also, game-theoretic approaches can be used to improve the system response. Kiennert \etal \cite{10.1145/3232848} survey strategies capable of analyzing the interactions between attackers and defenders, then responding to attacks, via game theory and Markov decision processes.

Based on how frequently the attacks occur, \textit{event-triggered control} schemes instead of time-triggered schemes emerged as appropriate tools to increase the resilience of control systems \cite{6425820, DIBAJI2019394}. The application of event-triggered control to the resilience of \gls*{cps} has been studied in \cite{Cetinkaya2016, YANG2017145, LEI2016286} where the triggering function to generate a new control input is based on the errors of the state variables. Sun \etal \cite{Sun__event_triggered_2022} propose an adaptive event-triggered resilient control to resist asynchronous data attack injection in industrial \gls*{cps} network communication. Their proposal uses a threshold that dynamically changes and adjusts the control strategy, according to the attack.

\textit{Advantages and limitations:} This approach provides a flexible and dynamic response mechanism that can work only when the system is under attack to provide graceful degradation. It may be designed to protect sensors, actuators, controllers of the network traffic. However, this approach may be hard to test and to ensure the stability of the control feedback when combining malicious and defensive actions over the physical process. It may have hidden undesirable actions or cascade effects that may be harmful to the system. In addition, it increases the complexity, making it complicated to test all possible combinations of malicious actions and dynamic defensive configurations.

\vspace{-.25cm}

\subsubsubsection{Programmable Networks} Some other proposals are based on programmable networking that enables efficient network configuration that can be used for neutralizing attacks. New networking functionality can be programmed using a minimal set of APIs (Application Programming Interfaces) to compose high-level services. This idea was proposed as a way to facilitate network evolution. Some solutions such as Open Signaling \cite{Campbell_opensignal}, Active Networking \cite{Tennenhouse_active_network}, and Netconf \cite{Netconf}, among others, are early programmable networking efforts and precursors to current technologies such as Software Defined Networking (SDN) \cite{Kreutz_SDN_survey}. In particular, SDN is a programmable networking paradigm in which the forwarding hardware is decoupled from control decisions. SDN proposes three different functionality planes: (1) data plane, (2) control plane, and (3) management plane. The data plane corresponds to the networking devices, which are responsible for forwarding the data. The control plane represents the protocols used to manage the data plane, such as, to populate the forwarding tables of the network devices. The management plane includes the high-level services and tools, used to remotely monitor and configure the control functionality. Security aspects may have an impact on different plans. For example, a network policy is defined in the management plane, then the control plane enforces the policy and the data plane executes it by forwarding data accordingly.

The idea of using programmable networks for improving security includes the management of denial-of-service (DoS) attacks~\cite{sahay} and segmentation of malicious traffic \cite{hadega,itl2018}. Programmable networks provide higher global visibility of the system, which is favorable for attack detection. In addition, a centralized control plane may allow further possibilities to achieve dynamic reconfiguration of network properties, e.g., application of countermeasures. Molina \etal \cite{molina_software-defined} survey approaches for SDN controllers that are able to establish different paths between sensors and actuators. Piedrahita \etal \cite{PIEDRAHITA2018} use SDN and network function virtualization to facilitate automatic incident response to a variety of attacks against industrial networks. The resources are assigned after an attack is detected. SDN and cloud-enabled virtual infrastructure help to respond automatically to sensor attacks and controller attacks by rerouting malicious traffic to a honeypot and transfer the services from the compromised device to a new virtualized device.

\textit{Advantages and limitations:} The programmable networks also provide a dynamic reconfiguration to respond at runtime to malicious actions in the network traffic. These approaches are flexible, however, it may be hard to analyze how the new network configuration affects the network delay and jitter, which is vital in real-time applications. Also, the reconfiguration increases the network complexity and the restoration work may induce hidden undesirable behaviors within the system.

\medskip

In this section, we have presented detection and reaction mechanisms for cyber-physical adversaries. However, despite the implemented mechanisms, it is still possible to have a system breach. For this reason, it is desired to implement cyber-resilience by design approaches to absorb, survive or recover from threats. Another cyber-resilience taxonomy can be found in \cite{book_resilience}. Cyber-resilience demands a system design that provides flexibility, adaptability, and agility to react in real-time to disturbances. In the next section, we survey techniques to build cyber-resilient systems.

\subsection{Cyber-Resilience Approaches}
\label{subsec:ch2_resilience_sota}

A growing number of technologies and architectural practices can be used to improve cyber-resilience. In the rest of this section, we cover techniques that may be used to build resilient systems. We provide a taxonomy of cyber-resilience techniques and a literature survey of different proposals that apply them.
We analyze the techniques according to the cyber-resilience phase they react and the \gls*{cps} layer they protect. A resilience solution may work in the absorb, survival, or recovery phase. The absorb phase limits the damage of the attack or extends the surface that the adversary has to attack to be successful. For example, by isolating resources, limiting adversary access, and changing or removing resources.
The survival phase objective is to maintain or maximize the duration of the correct function of the essential system mission. The recovery phase aims at transforming or reconstituting  the resources to recover the functionalities after the attack.  We also analyze at which level of the system design the resilience approach works. For example, it may be at the physical level considering the hardware of the components, at the control level to face adversaries that exploit the control theory mechanism that is running in the controllers, at the network or cyber level considering the communications or the software of the system. Table \ref{tab:techniques} sums up the different cyber-resilience strategies and scientific proposals that use them.

\subsubsection{\textbf{Architecture Design}}

These strategies involve modifying the system architecture to improve the resilience of the system to absorb or survive the attack impact \cite{9646342}.

\subsubsubsection{Diversity} It uses a heterogeneous set of technologies to minimize the impact of the attack. Different technologies will have different and independent vulnerabilities, which will make the adversary task harder to achieve. In addition, this technique  increases the adversary uncertainty and the resources required for a successful attack.

This technique can be applied, for example, using different hardware, software, firmware, or protocols \cite{9760016}. It is worth noting, that this technique requires adding new components. These components should be different from the previous ones because just adding redundancy makes the system still exploitable by the same adversaries using the same vulnerabilities as in the primary components.

When designing software diversification techniques, it is required to decide what to diversify and when to diversify it \cite{larsen_sok_2014}. To decide what to diversify, possible techniques are: (1) randomization which works as a compiler optimization and can be applied, for example, at the instruction level by substituting equivalent instruction or sequence of instructions; (2) randomizing the register allocation or reordering instruction. Another option is to apply this technique also at block, loops, functions, data, or even program levels. For example, at the functions level, it is possible to randomize the order of function parameters or the layout in the stack to prevent buffer overflow attacks. At the program level, similar strategies can be applied to randomize the order of the functions within executables and libraries. Different options to decide when to apply the diversification are at implementation time (i.e., when coding) \cite{532621}, at compiling and linking the source code \cite{COHEN1993565, Forrest_buildingdiverse, 10.1145/2508859.2516675, Jackson2011, 10.1007/978-3-642-00730-9_10, 10.1007/978-1-4614-5416-8_8, 10.1007/978-3-540-70542-0_1}, or at installation, loading, or execution time \cite{10.1145/948109.948146, 6494997, cispa450, 4768651}.

Other diversity solutions may work also in a detection-reaction manner. For example, Ouffoué \etal \cite{8029792, ouffoue:hal-03113828} use diversification to create attack-tolerant web services. They modeled the services to extract different implementations using variations in style, encoding, and language. The multiple services' implementations allow monitoring for attacks and react by changing the active implementation.

In the case of hardware diversification, it is required to design if all the different components will be active at the same time or if they will act as a cold backup that is activated after the primary system is attacked. For example, authors in \cite{chaves_improving_2017} use diversity to improve cyber-resilience for industrial control systems. The strategy is implemented using primary and redundant PLCs from different vendors to enhance cyber-resilience.

\textit{Advantages and limitations:} When diversity is used with different implementations, it helps to create a system with independent vulnerabilities. This way, when a component is attacked, it can be disabled to continue working with the diversified copy. This may be applied to sensors, actuators, and controllers attacks. The advantage is that the system keeps all the functionalities working and it is possible to ensure the correct behavior of the control. However, this approach may be expensive due to the required extra hardware and it only addresses attacks at the endpoints. Also, it requires extra management and maintenance effort, for example, to apply the software updates to a wider and more diverse group of components.

\vspace{-.25cm}

\subsubsubsection{Segmentation} The design of a \gls*{cps} must consider how to prevent attacks and be more tolerant to intrusions from the beginning. A network segmentation strategy separates logically or physically the components to reduce the attack surface. It also contains and limits the damages of a successful attack. The components may be separated based on their level of criticality, trustworthiness, or functionality \cite{gengeSegmentation, 10.1007/978-3-642-41488-6_12}.

According to the results achieved in  \cite{gengeSegmentation}, this technique also contributes to building more intrusion-tolerant \gls*{cps}. Network segmentation may be designed considering the Process-Aware Control approach presented in \cite{10.1007/978-3-642-41488-6_12}. It establishes that attacks on some components generate a greater risk than attacks on other components in the same system. For this reason, it is important to classify the different network components and the control loops according to the impact they may have on the operation of the \gls*{cps}. This approach would allow protecting the essential components in a better way. Following this idea, it also allows having the notion of \textit{more insecure} nodes (for example, a node that uses wireless communication technologies), e.g., to place them in a network segment 
separated from the other nodes that are considered as a trust zone.

A segmented architecture can help to absorb the impact of a compromise and prevent cascading failures \cite{9693217}. A network susceptible to large cascade failures is likely to have severe damage to disturbances, which limits the absorption and recovery required to build a resilient system. For this reason, the dependencies and links between nodes should be designed to minimize the likelihood that a failure propagates easily from one node to another.

\textit{Advantages and limitations:} This approach is easy to implement and effective to contain the consequences of a compromised component. It also limits cascade effects and the propagation of the attack within the network. The main limitation is that it does not help to recover the system to its normal behavior.

\subsubsection{\textbf{Reconfiguration}}

There are different possible reconfiguration options. This technique requires a situational awareness to select pre-considered options, ensuring the intended consequences. For example, in a denial-of-service (DoS) attack, we might dynamically over-provision additional processing capabilities. If an attack comes from the outside, we may reconfigure boundary protections and security policies. During a failure, we may shut down non-essential functions or initialize alternative capabilities to execute critical processing. We classify possible reconfiguration in the following categories.

% \medskip

\subsubsubsection{Isolation and Containment} These strategies aim at limiting the spread of the adversary by separating compromised from non-compromised components. For example, if an adversary controls a part of the system, it may be necessary to temporarily shut down it to close the adversary’s channel while critical mission functions are completed in another portion of the system.

Kwasinski in \cite{kwasinski_modeling_2020} analyzes this problem for power grid and he shows how service buffers, such as energy storage or a data connectivity reestablishment ensured time, help limit the impact of intra-dependencies on resilience. They explain that without service buffers, failures in an infrastructure component may immediately cascade within the system or onto other infrastructures. For this reason, resource buffers play a critical role in understanding cyber-physical interactions, limiting the negative effect of intra-dependencies, and improving resilience.

Xu \etal \cite{xu_islanding_2020} show that isolation and reconfiguration are effective approaches for service restoration and resilience enhancement. They propose a multi-stage switch strategy based on dynamic programming, considering both isolating and fault reconfiguration. First, they propose the construction of numerous expected fault scenarios. Then, they select some of them and develop their information entropy. Finally, for each typical scenario, a multi-stage switch strategy considers both isolation and fault reconfiguration, through dynamic programming.

Bellini \etal \cite{bellini_cyber_2019} analyze Internet of Things (IoT) resilience considering a network-based epidemic spreading approach. The mathematical model assesses infection and communication interactions to reduce a malware outbreak while maintaining the network functionalities at an acceptable level. Disconnecting a network region compromises connectivity. The mobility of resources to an affected area is of critical value for the immediate local control of outbreaks and  to prevent the spread.

Chen \etal \cite{chen_robustness_2020} analyze how attacks in communication networks may cause cascading failure in a physical power grid. They find that clusters in physical power grid and communication network are mutually interdependent to survive in cascading failure, operating in the form of isolated subsystems the failures remain interdependent to stay alive when cascading attacks occur. Hence, they consider survival clusters to adjust intra- and inter-links and study the robustness of the system in various attack scenes.

Haque \etal \cite{haque_modeling_2019} analyze resilience for energy delivery systems considering cyber components and service criticality. They estimate the criticality using graph Laplacian matrix and network performance after removing links (i.e., disabling control functions or services) and also analyze the cyber resilience by determining the critical devices using TOPSIS (Technique for Order Preference by Similarity to Ideal Solution) and AHP (Analytical Hierarchy Process) methods. They consider paths as a sequence of services or control functions and assume the removal of links as disabling the service or deactivating the control function rendered by the particular device.

\textit{Advantages and limitations:} As in the previous approach, this technique is effective in containing the consequences of an attack but it does not help to recover the system to its normal behavior. Isolating or disconnecting a component or part of the system in case of compromise prevents the spread and cascade failures. However, this might be also detrimental to the overall resilience of the system if the isolated component is needed to support other components that execute damage-absorbing actions. The recovery actions should be planned with this in mind.

\vspace{-.25cm}

\subsubsubsection{Dynamic Network Composition} This technique designs the system with dynamic capabilities to face the attacks. For example, distributing tasks in different organizations. Januario \etal \cite{januario_distributed_2019} propose a hierarchical multi-agent framework that is implemented over a distributed middleware with distributed physical devices. The architecture uses software-defined networks and cloud-based virtual infrastructures. Physical and cyber vulnerabilities are taken into account, and state and context awareness of the whole system are targeted. Each multi-agent executes a specific task and adapts its behavior depending on its location and environmental changes. In addition, Chen \etal \cite{chen_adaptive_2020} propose an approach to improve resilience using the synchronization of multi-agent systems that address faults and uncertainties in communication links. For that, they transform the resilient control problem into distributed state observers.

Marshall \etal \cite{marshall_context-driven_2019} present a context-driven decision engine for adaptive resilient control. It integrates diagnostic and prognostic heuristics to establish situational awareness and drive actions. The proposal assesses the system's state of health based on operational availability and drives control decisions based on scenario-specific constraints and priorities. Similarly, Ratasich \etal \cite{ratasich_self-healing_2017} presented a self-healing framework that uses structural adaptation, by adding and removing components, or by changing their interaction, at runtime. Segovia \etal \cite{segovia_reflective_2020} proposed an attenuation strategy that uses software-defined networks and software reflection. In case of attack, the approach creates dynamically in the network domain a component on the fly to help or assume the functions of the victim node.

\textit{Advantages and limitations:} This approach changes the configuration of the system periodically and increases the attack effort. In addition, the new configuration may force the adversary to re-implement the attack with each system change. This technique may be effective for attacks that compromise controllers of the network traffic. The limitation of this approach is that it may be hard to ensure the stability of the control feedback when combining malicious and defensive actions over the physical process. Also, it increases the complexity of the system and it is harder to test, manage and debug.

% \medskip

\subsubsubsection{Non-Persistence} This technique reduces the adversaries' opportunity to identify and exploit    vulnerabilities or maintain access over resources whose access is not continuous in time. It can be applied, for example, to data, applications, or connectivity, making them only accessible during a particular time. In addition, with this technique, a system can periodically refresh to a known previous image to ensure that the current image complies with a secure configuration.

Another option is to implement reversibility. This way, components are designed in a manner that allows them to revert to a safe mode when failed or compromised. This means that the component in the failed mode should not cause any further harm to other components in the system; and second, it should be possible to reverse the state of the component in the process of recovering the system. The system can periodically refresh to a previously known image to ensure that the current system image is correct.

For example, Griffioen \etal \cite{9147232} present a decentralized control system and a procedure to determine when agents should communicate with one another after having been disconnected from the network for a period of time. When agents communicate with one another, they guarantee system resilience against malicious adversaries by utilizing software rejuvenation, a prevention mechanism against unanticipated and undetectable attacks on cyber-physical systems. Without implementing any detection algorithm, the system is periodically refreshed with a secure and trusted copy of the control software to eliminate any malicious modifications to the run-time code and data that may have corrupted the controller.

Pradhan \etal \cite{PRADHAN2016344} present a runtime infrastructure that provides autonomous resilience via self-reconfiguration. The approach relies on the implicit encoding of all possible states a system can reach (the configuration space) and it consists of relevant information about different system goals, functionalities, services, resources, and constraints. At any given time, there is exactly one configuration point that represents the current state of a platform. At runtime, when a configuration point is deemed faulty, the self-reconfiguration infrastructure computes a valid new configuration point that belongs to the same configuration space, and then transition, migrate, or reconfigure to the newly computed configuration point such that failures or anomalies are mitigated.

\textit{Advantages and limitations:} This approach returns the system to a previously safe and known state, which ensures the correct behavior. It is effective for attacks that compromise specific devices such as sensors, actuators, controllers, routers, or switches. However, this solution does not last long, due to the vulnerabilities exploited by the adversary are still present in the previous image, and they can be exploited again.

\subsubsection{\textbf{Moving Target Defenses}}

A static structure allows adversaries to collect information and perform long-term analysis. In addition, the uniformity of components allows adversaries to expand the damage scope after they find one vulnerability. For this reason,
Moving Target Defense (MTD) approaches provide strategies that change the system over time to increase its complexity, attack cost, or limit the exposure of vulnerabilities \cite{zheng_survey_2019}. The mechanisms are usually applied at the network or the node level~\cite{lei_moving_2018}. Next, we summarize  proposals for both levels as well as approaches specially designed for \gls*{cps}.

% \medskip

\subsubsubsection{Network MTD Approaches} The \textit{endpoint information} (such as MAC address, IP address, port, protocol, or encryption algorithm) and the \textit{forwarding path} (links and routing nodes) are two key elements in network transmission and it can be used to identify the source and destination nodes. Hence, it is important to protect this information as part of the attack surface.

Some approaches that protect the endpoint information are as follows.
Antonatos \etal \cite{ANTONATOS20073471} propose the use of Network Address Space Randomization (NASR) to handle worm attacks. The method analyzes and discriminates the potentially infected endpoints and the nodes are forced to frequently change their IP address by using DHCP protocol.
Al-Shaer \etal \cite{10.1007} proposed Random Host Mutation that assigns virtual IP addresses that change randomly and synchronously in a distributed way over time. To prevent disruption of active connections, the IP address mutation is managed by network appliances and transparent to the end host.

MacFarland \etal \cite{Macfarl_thesdn} hide the endpoint MAC, IP, and port numbers by setting up a DNS hopping controller and synthetic addressing information in place of the real one with the help of NAT rules. This %synthetic information
can be considered to be chosen at random within certain validity constraints.

Other approaches protect the forwarding path information, i.e., it randomly selects routing nodes to change the forwarding paths while ensuring reachability. For example, Dolev \etal \cite{6924217} use a secret sharing technique to encrypt  data and create \textit{n} shares, and only fewer than \textit{k} parts can be allowed to transmit in the same path. In addition, to reconstruct the data, the destination needs to have at least \textit{k} shares out of the \textit{n} shares that were sent. The approach objective is to provide private and secure interconnection between the data centers.
Aseeri \etal \cite{Aseeri} propose an approach to improve the diversity of forwarding paths to deal with eavesdropping attacks in the SDN data plane. It uses bidirectional multiple routing paths to reduce the severity of data leakage. The SDN controller applies the multipath mechanism both ways, from the sender side and the receiver side. By negotiating migrating paths between source and destination, the forwarding path is changed randomly during transmission.

Duan \etal \cite{6682715} propose a Random Route Mutation technique that enables changing randomly the route of the multiple flows in a network simultaneously to defend against reconnaissance, eavesdropping, and DoS attacks while preserving end-to-end QoS properties.
Ma \etal \cite{978-3-319-50011} propose an approach for self-adaptive end-point hopping, which is based on adversary strategy awareness and implemented using SDN. This method periodically changes the network configuration in use by communicating endpoints. \CH{Potteiger \etal \cite{Bradley_Potteiger_2022} propose to implement MTD techniques such as Address Space Randomization (ASR), and Data Space Randomization (DSR) in a mixed time and event-triggered architecture to maintain the safety and availability during the attack. Mixing both architectures allows the system to support predictable operation during normal circumstances while maintaining rapid detection and reconfiguration during an attack. Xu \etal \cite{Xiaoyu_moving_2022} propose an MTD technique with a  routing randomization method based on deep reinforcement learning. This proposal improves the security against eavesdropping attacks, improving the random routing granularity, real-time and accurate network state awareness, and powerful decision-making. Azab \etal \cite{Azab_moving_2022} propose a novel MTD approach using multi-controller management of SDN. The objective of this multi-controller approach is to detour the runtime workload among multiple controllers and control misbehavior detection without impact on \gls*{cps} performance.}

\textit{Advantages and limitations:} This approach is similar to \textit{Programmable Networks}, the difference is that the re-configurations are periodicals and not triggered by any detection. Due to the pre-configurated system change, it may be possible to predict better the response of the system in each change and also in case of an attack. The approach is effective for attacks that compromise network traffic. One limitation is that the re-configurations increase the network complexity impacting negatively the debugging and managing effort. It may also impact the network performance in each reconfiguration period. For instance, creating loops until all the paths 
are updated or affecting the latency between nodes.
% \medskip

\subsubsubsection{Node MTD Approaches} Platform environment and software applications can be diversified to protect from adversaries. Diversity proposes to have many forms of the same object because this design can reduce the probability of intrusion \cite{verissimo_intrusion-tolerant_2003}. Address space, instructions, or data randomization are three typical ways to achieve platform environment diversification \cite{Forrest}. Another technique is software application isomerization. In software engineering, isomerization is a mechanism that changes codes dynamically to enhance the heterogeneity of software applications under the premise of ensuring functional equivalence. Depending on the application software life cycle, it can be divided into transformation mechanisms adopted during software compilation and link or transforming mechanisms implemented during software load and execution \cite{lei_moving_2018}. In addition, programmable reflection is a meta-programming technique that has the potential to allow a programmable system to manipulate itself at runtime~\cite{reflect_Ana}.

The previous techniques are software techniques that can be applied to a wide variety of systems. Some CPS-specific MTD approaches have been proposed to control adversaries situated in the end devices, i.e., actuators and sensors. For example, in~\cite{giraldo_moving_2019}, Giraldo \etal propose an MTD strategy that randomly changes the availability of the sensor data, so that it is harder for adversaries to achieve stealthy attacks. This approach uses switched control systems that allow detecting sensor compromise and minimizing the impact of false-data injection attacks. \CH{In \cite{giraldo_moving_2022}, Giraldo \etal present a novel approach for MTD using IoT-enable Data Replication (MTD-IDR). They utilize liner-matrix inequities for the optimization problem, in order to select and optimize the number of replicas of each communicated signal in the system. This approach prevents stealthy attacks and reduces the accuracy of attack to learn the system's model, nevertheless the energy consumption increases and the bandwidth is reduced.}
Griffioen \etal \cite{griffioen_moving_2019} propose an MTD approach for recognizing and isolating \gls*{cps} integrity attacks on a set of sensors and actuators by introducing stochastic time-varying parameters in the control system. The underlying random dynamics of the system limit the adversary's knowledge of the model. \CH{Liu \etal \cite{Liu_moving_2021} propose a strategy by proactively perturbing the primary control gains of the power converter device in DC microgrids (DCmGs) to defend against deception attacks. They highlight the importance of providing explicit conditions for the magnitude and the frequency of the perturbation in order to ensure the voltage stability of the system.}

Weerakkody \etal \cite{weerakkody_moving_2016} propose an MTD approach to minimize identification in \gls*{cps}, i.e., to limit the adversary's knowledge of the system model to identify sensor attacks by changing the dynamics of the system as a function of time.
Kanellopoulos \etal \cite{kanellopoulos_moving_2019} propose an approach to mitigate sensor and actuator attacks by formulating a control algorithm based on MTD that provides a proactive and reactive defense mechanism. It uses a stochastic switching structure to alter the parameters of the system and make it more difficult for the adversary to perform a system reconnaissance. Segovia \etal \cite{9266030} propose an MTD approach that changes the \gls*{cps} physical model that executes in each node periodically. The system is modeled as a switched control system to improve resilience.

\textit{Advantages and limitations:} Similarly to the previous approach, the re-configurations are periodicals and not triggered by any detection. This is an advantage because modeling the system as a Switched Control System makes it possible to better predict the stability of the system in each change to ensure its correct behavior. In this case, the control theory provides strong mathematical models to understand, limit the damage, and predict the system behavior under attack. This approach may be efficient for sensor, actuator, controller, or network attacks depending on how it is designed. One limitation is that the re-configurations increase the system complexity making the debugging and testing effort bigger.

\subsubsection{\textbf{Dynamic Software Evolution}}
Dynamic software evolution uses code generation or modification at runtime to adapt the system behavior and face adversaries.  We can differentiate two main approaches: \textit{Runtime Code Generation} and \textit{Software Reflection}.

The former, Runtime Code Generation, is a particular case of code generation techniques used to create source code at runtime. Some languages support this feature, for example, .NET which provides a mechanism that produces source code in multiple programming languages at runtime, based on a single model that represents the code to render in a language-independent object model. This way, programs can be dynamically created, compiled, and executed at runtime.	Code generation involves creating code that never has to be modified once it is generated. If a problem arises, the problem should be fixed in the code generator, and not in the generated source files. This technique may be used to generate diversity in the created software.

The latter, Software Reflection or Self-Modifying Code, is another technique that allows a system to adapt itself through the ability to examine and modify its execution behavior at runtime. As a mitigation technique, software reflection has the potential to allow a system to react and defend itself against availability threats. When malicious activity is detected, the system shall dynamically change the implementation to activate remediation techniques to guarantee that the system will continue to work. Software reflection provides the ability to analyze, inspect, and modify the structure and behavior of an application at runtime. This allows the code to inspect other codes within the same system or even itself. Reflection allows inspecting classes, examining fields, changing accessibility flags, dynamic class loading, method invocation, and attribute usage at runtime even if that information is unavailable at compile time. Also, it is possible to use data marshaling and pull data from an outside source and load it into a Java object or use reflection to execute it.

He \etal \cite{he_software_2008} propose an approach to modify the software runtime architecture through meta-operators based on reflection. Similarly, Kon \etal \cite{kon_case_2002} propose a reflective middleware to deal with highly dynamic environments, supporting the development of flexible and adaptive systems and applications. Mavrogiannopoulos \etal \cite{10.1016/j.cose.2011.08.007} present a taxonomy of self-modifying code with the purpose of obfuscation.

\textit{Advantages and limitations:} This approach is flexible and dynamic and it works for attacks on sensors, actuators, and controllers. However, it is harder to have control over what is being executed in each node and its effects on the system stability. Also, due to the difficulty of understanding what is being executed, it may be harder to test, debug, and protect the system.

\subsubsection{\textbf{Consensus, Secret Sharing, and Distributed Trust}}

Both consensus and distributed trust approaches have been largely investigated for general computer science problems where some of the subsystems are untrustworthy.

Consensus protocols provide resilience to the byzantine problem, i.e., in the presence of malicious nodes that send incorrect messages to deceive the system. These consensus approaches may be applied at the network level which has been largely studied by the distributed computing research community \cite{Lamport_byzantine, Fekete_consensus, leblanc_resilient}, or it may also be applied at the control level which is an active research area in the control theory community. In this case, at each update, the controller ignores suspicious values and computes the control input with the non-suspicious values. For example, using Distributed Kalman Filter for resilient state estimation \cite{mahmoud_distributed_2013, wen_distributed_2018} or other distributed observers strategies to manage sensor compromise \cite{severson_resilient_2020, 8270595}.
Other strategies are distributed function calculation in the presence of malicious agents \cite{5605238}, distributed multi-agent consensus \cite{amini_performance_2019, saldana_resilient_2017, meng_studies_2014, 7906513, DIBAJI2017123}, resilient vector consensus \cite{yan_resilient_2020, shabbir_resilient_2020}, and resilient leader-followers consensus approaches \cite{usevitch_resilient_2019, zegers_event-triggered_2019}.

Techniques such as secret sharing schemes \cite{Shamir_secret, Brickell_secret, beimel_secret} and distributed trust \cite{Abdul_distribTrust, Josang_distribTrust} may be used to implement, for example, mechanisms that divide the control into shares, such that the system needs to reach a given threshold before granting control, \ie, a data $D$ is divided into $n$ pieces in such a way that $D$ is easily reconstructable from any k pieces, but even complete knowledge of $k-1$ pieces reveals no information about $D$.
Secret-sharing schemes are important tools in cryptography used in many security problems such as multiparty computation, Byzantine agreement, threshold cryptography, access control, attribute-based encryption, distributed certificate authorities, distributed information storage, key management in ad-hoc networks, electronic voting, and many others.
A classical approach to building secret-sharing schemes is Shamir’s threshold approach \cite{Shamir_secret}, which divides the data $D$ using a polynomial of grade $n$. The correctness and privacy of this scheme follow from the Lagrange’s interpolation theorem. The undirected s-t-connectivity approach \cite{beimel_secret} builds the scheme using an undirected graph structure whose share parties between entities are mapped to edges, nodes, and paths to connect those nodes. Other existing schemes are based on monotone formulas, for example, the proposal in Ito \etal~\cite{Ito_4430720906}, the monotone formulas construction \cite{10.1007/0-387-34799-2_3}, and the monotone span programs construction \cite{10.1007/3-540-46885-4_45, 336536}. A monotone function is a function entirely non-increasing or non-decreasing, \ie, its first derivative does not change the sign. Every monotone formula computes a monotone function and every monotone function can be implemented using just AND and OR operators. 
Benaloh and Leichter \cite{10.1007/0-387-34799-2_3} proved that if an access structure can be described by a monotone formula then it has an efficient perfect secret-sharing scheme.

The distributed trust aims at interacting with the most secure, honest, and trustworthy entities because this minimizes the exposure to risky transactions.
One strategy for distributed trust is a human-like mechanism based on reputation that chooses between benevolent and malicious behavior. Then using relationships and inferring rules, different levels of trust are derived for other entities \cite{Josang_distribTrust}. This way, reputation is an assessment based on the history of interactions with or observations of an entity, either directly with the evaluator (personal experience) or as reported by others (recommendations or third-party verification). A second mechanism to determine trust is using policies that describe the conditions necessary to obtain trust. It can also prescribe actions and outcomes if certain conditions are met \cite{ARTZ200758}. Policies frequently involve the exchange or verification of credentials, which are information issued (and sometimes endorsed using a digital signature) by one entity, and may describe qualities or features of another entity. Also, Distributed Ledger Technologies, like Blockchain, are characterized by transparency, traceability, and security by design. These features make the adoption of Blockchain attractive to enhance information security, privacy, and trustworthiness in very different contexts including distributed trust \cite{8970496}.

\textit{Advantages and limitations:} This approach is useful to address compromised components and it is effective for a determined number of compromised devices. As a result, the information used for the feedback control is more accurate and it is harder to execute commands based on fake information. However, when the majority is wrong, the stability and the correct behavior of the system are also compromised. Another limitation is the required time to synchronize the information between all the nodes. For this reason, the decision process can take a long time which is not suitable for real-time applications.

\subsubsection{\textbf{Game Theory}}

Approaches based on game-theoretic strategies use mathematical models to analyze the situation where players choose a different action in an attempt to maximize their returns \cite{ILAVENDHAN201846}. It studies decisions made in an environment in which multiple players interact with each other in a strategic setup. This means that game-theoretic approaches provide resilience trying to maximize the cost of attacking the system or minimize the damage that an adversary can apply to the system. For that, each player tries to optimize an objective function. This objective function depends on the choices of the other players in the game. Thus, players cannot optimize their objective independently of the choices of other players.
This technique has been proposed to respond to attacks where the defender chooses the optimal response according to the adversarial actions. Game theory provides tools to model advanced adversaries who know the defense strategies and can adjust the attack strategies accordingly. In addition, it is possible to define games in both physical and cyber layers.

In the last years, there have been many proposals on game-theoretic approaches for \gls*{cps}. For example, Huang et Zhu \cite{huang_dynamic_2020} propose a dynamic game for long-term interaction between a stealthy adversary and a proactive defender. The stealthy and deceptive behaviors are captured by the multi-stage game of incomplete information, where each player has his private information unknown to the other. Both players act strategically according to their beliefs which are formed by multi-stage observation and learning. In addition, Hasan \etal \cite{hasan_game-theoretic_2020} design an adversary-defender game-theoretic model for power systems. The adversary can identify the chronological order in which the critical substations and their protection assemblies can be attacked to maximize the overall system damage. The defender can intelligently identify the critical substations to protect such that the system damage can be minimized. Ismail \etal \cite{10.1007/978-3-319-47413-7_10} model the interactions between an attacker and a defender and derive the minimum defense resources required and the optimal strategy of the defender that minimizes the risk. The solution is analyzed in power systems. Also, Rao \etal \cite{rao_resilience_2015} propose a resilience approach using a game approach to face adversaries. Their functions consist of an infrastructure survival probability and a cost expressed in terms of the number of components attacked and reinforced. Zhu and Basar \cite{zhu_game-theoretic_2013} propose a game-theoretic approach to manipulate the attack surface of the network and create a moving target defense. The notion of attack surface is defined as the set of vulnerabilities of the system that can potentially be exploited by the adversary. The essential goal is to find an optimal configuration policy for the defender to shift the attack surface that minimizes its risk and damage.

Game-theoretic approaches have also been proposed to learn adversary models and estimate their knowledge about the system dynamics. For example, Sanjab and Saad \cite{sanjab_bounded_2016} propose a game-theoretic approach to analyze the interactions between one defender and one adversary over a \gls*{cps}. In this game, the adversary launches cyber-attacks on several cyber components of the \gls*{cps} to maximize the potential harm to the physical system while the system chooses to defend a set of cyber nodes to thwart the attacks and minimize potential damage to the physical side. Similarly, Kanellopoulos and Vamvoudakis \cite{kanellopoulos_non-equilibrium_2019} consider the problem of identifying the cognitive capabilities of adversaries. To categorize them, they use an iterative method of optimal responses that determine the policy of an agent with a determined level of intelligence. Then, they formulate a learning algorithm to train the different intelligence levels without any knowledge about the physics of the system.

\textit{Advantages and limitations:} This approach provides a quantitative mechanism for deciding the optimal strategy to face an attack. It may be effective for attacks on sensors, actuators, controllers, or network traffic depending on how the approach is designed. However, most of the existing proposals focus on cyber or network aspects, without considering the physical model of the process. In addition, as the decisions are calculated at runtime it may be hard to analyze and predict the stability of the physical process.

\section{Discussion}
\label{sec:ch2_why_CT}

Control theory and cybersecurity are research areas that provide significant contributions to solve security issues in \gls*{cps} from different perspectives. Similarly to the IoT domain~\cite{10.1145/3462513}, resilience in the \gls*{cps} domain is a dual problem with a part in the cyber world and the other part in the physical one. As pointed out in \cite{sanchez_bibliographical_2019, 6580348},  both such domains are complementary disciplines that working together have the potential to provide more efficient and effective solutions.

Control theory provides models that precisely describe the underlying physical process, which enables the prediction of future behavior and unforeseen deviations from it. It models the system to analyze attacks and their corresponding detection, mitigation, and recovery schemes. The cybersecurity research community also offers different approaches for numerous security problems in \gls*{cps}. Such approaches typically focus on the cyber aspects, such as communication networks, protocols, software, and data.

According to \cite{6580348}, \gls*{cps} security can be divided into two main categories: information security which focuses on cyber and data security, provides methods that are effective on software layers without using any physical model; and secure control theory, which studies how cyber-attacks affect the control system’s physical dynamics. Ensuring safety using only information security tools is not sufficient for \gls*{cps}. Therefore, they should be complemented with secure control theory that provides an attack model and a description of the interaction between the physical world and the control system. It provides a better understanding of the attacks' consequences, and the development of new detection methods, algorithms, and architectures, that make the control systems more resilient to possible attacks and failures.

Certain attacks are undetectable by traditional control-theoretic approaches, for example in situations when the adversary modifies inputs and outputs to be correlated with the estimated model or when the values are chosen by the adversary to fulfill certain properties as described in \cite{teixeira2012attack,Teixeira2015}. The incorporation of cybersecurity strategies to control theory approaches, provided new tools to build approaches to solve this issue as explained in Section \ref{sec:ch2_techniques}. Moreover, cybersecurity approaches do not cover all the possible vulnerabilities in the cyber components. Mechanisms to protect specific vulnerabilities may not exist or be too expensive to implement, and even when they are implemented they are also not free of false negatives.

Furthermore, due to the strong coupling between cyber and physical domains, the tools and methodologies developed to ensure cybersecurity are insufficient to secure \gls*{cps}. For instance, they can fail against purely physical attacks. As an example \cite{weerakkody_resilient_2020}, the confidentiality of encrypted sensor measurements can be violated by placing unencrypted malicious sensors in close proximity to encrypted sensors. The integrity of sensor measurements can be modified by changing a sensor’s local environment while control inputs can be changed by directly manipulating system actuators. In such a scenario, message authentication codes or digital signatures fail to recognize an attack. Availability can be compromised by physically shielding sensors and actuators. In this case, anti-jamming and denial-of-service techniques will fail.

The large scale of a \gls*{cps} may turn physical protection impractical, leaving the system vulnerable to the previous examples. However, in addition to the exposed vulnerabilities created by basic physical attacks, it is possible to create more advanced cyber-physical attacks that generate the same physical effects but using a remote connection and injecting malicious traffic. As shown in Section \ref{sec:CPS_attacks}, malicious traffic can be confused with legitimate traffic and be undetectable. This way, by using control theory models, it is possible to implement new advanced and coordinated attacks to exploit \gls*{cps}. These attacks are capable of bypassing cyber detection as discussed in the literature: the false data injection attack \cite{Mo2010FalseDI, mo2010false}, the replay attack \cite{Mo_2014}, the zero-dynamics attack \cite{teixeira2012revealing}, and the covert attack \cite{smith2011decoupled}. Last but not least, insider adversaries and human error that generate security breaches have to be also considered to ensure safety.

\section{Open Challenges}
\label{sec:ch6_future_work}

The limitations highlighted in the previous section open several guidelines for future research work on the subjects surveyed in this article that would be beneficial for wider adoption of resilience methods and techniques for \gls*{cps}. Some representative guidelines are briefly presented next, in this section.

\vspace{-.35cm}

\subsection{System Modeling}

In terms of modeling, a \emph{higher interaction between system components}, e.g., cyber and physical components, would make the results more consistent and convincing. Indeed, a proper combination of the cyber-network and control-physical layers could be expanded towards next-generation cyber-physical systems able to properly correlate and repair cross-layer security incidents.
Most of the existing resilience techniques and measures focus on protecting the network, software, or physical components in an independent manner. In a \gls*{cps} these elements work together and coordinated actions to attack vulnerabilities in the different components that may have dangerous consequences.
More integration between the different layers creates systems with better capabilities to react and defend from adversaries.
For that reason, resilience techniques should integrate these concepts and have a global view of the components and their interaction because approaching the problem with partial and independent views is not enough to solve the existing security issues.

Concerning \emph{resilient control and attack models}, the control theory domain shows to be more mature than the computer science and cybersecurity fields. However, the integration of both domains creates new challenges that need to be addressed. For example, how to create attack-tolerant control, \ie, how to design robust control that considers possible attacks. Proactive algorithms and system architectures that are robust to attacks, ensure stability, and the performance thresholds are still required. In addition, the state of the cyber and network components should also be taken into account to consider factors such as the nodes' states and quality of service. To achieve that, it is also needed to improve the existing attack models, \ie, create attack models that characterize better the capabilities of the adversaries. One adversary model was developed in \cite{Teixeira2015} which is based on the available resources to an adversary. However, better models are still required including information such as their computational power, the type of access they may have, the data they collect, their collaborative capabilities, and signals an adversary has access to. This information helps to understand the logic behind the associated defense mechanisms, e.g., to improve or compare them with other security mechanisms.

A promising research opportunity related to the topics of this article can be explored around the use of \emph{digital twins}. In Section \ref{sec:ch2_CPS}, we present how to design the control loops using Kalman filters as estimators used for stochastic cases.
Such filters explicitly use a noise model for both state and output processes considering the stochastic nature of the dynamical system. Thus, it is more appropriate for \gls*{cps} and, in general, performs better for stochastic systems. Conversely, an observer, such as the Luenberger observer \cite{luenberger_introduction_1971}, is typically restricted to the deterministic cases, \ie, when there is no randomness in the states. Observers are used to estimate unmeasured states of a system and have been proposed to detect attacks in \gls*{cps}. The principle of estimators and observers are similar. An observer is a continuous-time dynamical system that takes as input the measured input and measured output of the system, and produces an estimate of the state of the system as output.

\vspace{-.35cm}

\subsection{Metrics and Evaluation Methods}

Some more efforts are needed on specifying \emph{complexity management to anticipate impacts on resilience}, e.g., to evaluate in a resilience approach how to manage the complexity of the proposal and how to anticipate the impact it may have on the system resilience. The resilience of a system is influenced by several factors that can be managed or exploited to enhance resilience \cite{linkov_fundamental_2019, book_resilience}. All resilience-enhancing measures can also cause a negative effect leading to an overall reduction in resilience. For example, to improve resilience, it may be required to use more complexity, such as using new connections, new components, more diversity, etc. As the number and heterogeneity of components grow, they offer more opportunities to regenerate the system. Agents may be able to use additional links to different elements or find replacement resources to ultimately restore their functions. However, high complexity may lead to interactions that are hard to understand, analyze, and protect, causing unforeseen side effects. As a result, greater complexity may also reduce the resiliency of the system.  Another example is fail-safe designs that disconnect a component or part of the system in case of compromise. This action prevents the spread and cascade failures. However, this might be detrimental to the overall resilience of the system if the component is needed to support other components that execute damage-absorbing actions.

The increase in complexity may lead to lower resilience by increasing the number of ways in which one failed component may cause the failure of another. Therefore, in most cases, greater complexity should be avoided when possible unless it directly supports resilience functions.
As a consequence, it is not enough only an analysis of the performance impact of an approach. Resilience proposals may also have hidden impacts on the system's behavior and complexity that should be evaluated to consider the reduction in the overall resilience. The quantification and evaluation of this aspect is not trivial. An approach should never be implemented in production systems without an appropriate evaluation of these factors. How to appropriately analyze and measure the resilience enhancement to reveal potential negative impacts and systemic effects is another future research work.

Another line in terms of metrics and evaluation methods relies on \emph{safety ensuring and testing automation}. \gls*{cps} normally provide critical functionalities. It is essential to ensure stability and correct behavior even under an attack when the inputs are specially modified for malicious purposes. In addition, triggering defensive actions increases the complexity of the system. Hence, with all these aspects happening at the same time may be hard to ensure that safety-critical functions will continue to work properly in any context or situation. Testing and validating the security proposals to ensure physical safety is still an open issue.

\subsection{Testing and Validation Environments}

There is a need to develop global approaches in terms of \emph{scalability validation}. Indeed, real \gls*{cps} may scale into networks with hundreds or thousands of devices. Conti \etal \cite{conti2021survey} survey validation testbeds and datasets for \gls*{cps}.  As a result, we can observe that scalability makes it difficult to test the system in an integrated manner considering physical, network, and cyber components. To test scalability normally simulation tools are used, but they abstract or forget the physical process part which is the essential part of the \gls*{cps}. The ideal validation option is experimental testbeds, which may be expensive and also exists limited stable testbed scenarios. Thus, testing scalability while combining physical process, network, and software components is still a challenge. We highlight the need for better \gls*{cps} testing and validation environments. Numeric simulation tools, such as Matlab\textsuperscript{\textregistered} and Simulink\textsuperscript{\textregistered}, do not integrate the network and cyber aspects. Network simulation tools are conceived for traditional IT systems and do not integrate the physical process. Hence, performance validation in simulation platforms only gives a partial overview of the whole problem.

In particular, for testing network aspects, it will not be enough to test with reduced quantities of the devices. This presents two new issues. First, creating such a testbed is not easy due to the required investment.  Second, the existing testbed scenarios consider only a limited quantity of devices. 
The lack of realistic scenarios is mainly due to the complexity of creating system models describing the different aspects of a physical process, such as the existing physical process reactions, the physical model involved in those reactions, the physical equipment or components required, the safety and operating constraints, the operating cost function, the sensor signal noise, the process randomness, among others \cite{krotofil2015rocking}. Designing such a system is a huge effort and insights into real industrial systems are not possible due to justified confidentiality issues.

\section{Conclusion}
\label{sec:conclusion}

\CH{In Cyber-Physical Systems (CPS), adversaries may disrupt physical processes by injecting malicious traffic, e.g., cyber-physical attacks may use coordinated cross-layer techniques, to get control over the cyber or network layers and disrupt the physical devices. For this reason, attacks over critical processes may end up affecting people, physical environments, and companies. To develop comprehensive protection for \gls*{cps}, it is required to layer the three following protection mechanisms: prevention to postpone the attack as much as possible, detection-reaction to identify the attacks and attenuate them, and cyber-resilience to contain the impact of the attack while providing essential services and restoring normal operations as soon as possible.}

\CH{Cyber-resilience is essential for critical systems that monitor industrial and complex infrastructures based on networked control systems \cite{10.1007/978-3-319-76687-4_3}. If the defense strategy relies only on detection and reaction approaches, the system is not protected in case of false negatives, i.e., undetectable attacks or extremely rare events that are not considered in risk management. Attacks might also come from inside, for example, from highly skilled employees acting as malicious insiders. The knowledge that such insiders possess about the system gives them unrestricted access to steal or modify data or even deactivate critical functionalities. It is important to have a \gls*{cps} capable of maintaining the stability of the system during such situations. The system should be protected at all times including the time required for detecting and responding to attacks. Otherwise, the system could experience disruption, leading to damages.}

\CH{In this article, we presented a systematization of knowledge about existing scientific efforts of making \gls*{cps} cyber-resilient. We systematically surveyed recent literature addressing the topic, with a specific focus on techniques that may be used on \gls*{cps}.  We started by surveying control theory formalities for \gls*{cps} and cyber-physical attacks. }
Then, we analyzed detection and mitigation techniques to protect \gls*{cps}. We surveyed some current trends in terms of detection based on control-theoretic model-based approaches that incorporate the physical model to detect cyber-physical adversaries. We also surveyed
mitigation techniques aiming to optimize the recovery response of a system under attack. The proposals to build cyber-resilient systems turn around techniques such as diversity, segmentation, resilient control, system reconfiguration, dynamic software evolution, moving target defense, consensus, and game theory paradigms. These techniques provide the ability to absorb, survive, or recover from an attack. 

We discussed how the techniques have evolved and we brought clarity to this complex field by treating the major axes of resilience techniques. We identified that the difference between the detection-reaction paradigm and resilience is not clearly defined in the literature, and often, the two concepts are confused. This problem arises for different causes. Firstly, because resilient designs are not easy to conceive. Our natural way of reasoning about security instructions is to detect the problem and then react. Another reason is probably that control theory and computer science have different definitions for the resilience concept. Control theory calls resilient a controller that can keep an understanding of the system state and calculate correct control signals despite malicious information injected at any point of the control loop. To achieve this, the control theory community normally uses approaches that in computer science are considered detection-reaction approaches. On the other hand, from a computer science perspective, a resilient system is capable to prepare, absorb, recover, and adapt to adverse effects. Or as we prefer to define it, a resilient system is capable of maintaining the core set of critical functionalities despite ongoing adversarial misbehavior and guaranteeing the recovery of the normal operation within a predefined cost limit.

\CH{As a result of the literature analysis, we identified plenty of research efforts in terms of detection techniques and state estimation to maintain an awareness of the system state despite an attack. However, much less efforts exist in terms of remediation approaches to attenuate the attacks. We identified a lack of adapted resilience techniques for the \gls*{cps} particular needs. The research in resilience for \gls*{cps} can be extended and we pointed out several promising directions for future work, with a focus on the practical aspects of cyber-resilience, such as the use of metrics and evaluation methods, as well as testing and validation environments.}

\medskip

%\begin{scriptsize}
\noindent \textbf{Acknowledgements ---} \CH{The authors thank the anonymous referees for their valuable comments and helpful suggestions. The authors also acknowledge support from the Cyber CNI chair of the Institut Mines-T\'el\'ecom, as well as support from the European Commission, under grant agreement 830892 (H2020 SPARTA project).}
% \end{scriptsize}

\bibliographystyle{unsrt_MS}
\bibliography{my_thesis}

\end{document}